\documentclass[aps, prx, reprint, showpacs, superscriptaddress]{revtex4-2}
\usepackage{graphicx}
\graphicspath{{imag/}}
\usepackage{amsmath, bm, amsfonts, bbm}
\usepackage{hyperref}%
\usepackage{tikz}

\newcommand{\rd}{\textrm{d}}
\newcommand{\re}{\textrm{e}}
\newcommand{\inp}[2]{\langle#1|#2\rangle}
\newcommand{\ket}[1]{|#1\rangle}
\newcommand{\bra}[1]{\langle#1|}
\newcommand{\ave}[1]{\langle #1 \rangle}
\makeatletter
\newcommand*{\rom}[1]{\expandafter\@slowromancap\romannumeral #1@}
\makeatother

\begin{document}
\title{Electrical Breakdown of Excitonic Insulators}

\author{Yuelin Shao}
\email{ylshao@iphy.ac.cn}
\affiliation{Beijing National Laboratory for Condensed Matter Physics and Institute of Physics, Chinese Academy of Sciences,
Beijing 100190, China}
\affiliation{School of Physical Sciences, University of Chinese Academy of Sciences, Beijing 100049, China}
\author{Xi Dai}
\email{daix@ust.hk}
\affiliation{Department of Physics, The Hongkong University of Science and Technology,
Clear Water Bay, Kowloon 999077, Hong Kong, China}

\date{\today}

\begin{abstract}


In this paper, we propose a new electrical breakdown mechanism for exciton insulators in the BCS limit, which differs fundamentally from the Zener breakdown mechanism observed in traditional band insulators. 
Our new mechanism results from the instability of the many-body ground state for exciton condensation, caused by the strong competition between the polarization and condensation energies in the presence of an electric field. 
We refer to this mechanism as ``many-body breakdown''. 
To investigate this new mechanism, we propose a BCS-type trial wave function under finite electric fields and use it to study the many-body breakdown numerically. 
Our results reveal two different types of electric breakdown behavior. 
If the system size is larger than a critical value, the Zener tunneling process is first turned on when an electrical field is applied, but the excitonic gap remains until the field strength reaches the critical value of the many-body breakdown, after which the excitonic gap disappears and the system becomes a highly conductive metallic state. 
However, if the system size is much smaller than the critical value, the intermediate tunneling phase disappears since the many-body breakdown happens before the onset of Zener tunneling. 
The sudden disappearance of the local gap leads to an ``off-on'' feature in the current-voltage ($I-V$) curve, providing a straightforward way to distinguish excitonic insulators from normal insulators.

\end{abstract}
\pacs{}
\maketitle

\section{Introduction}

The excitonic insulator is an insulating phase where
electron-hole pairs condensate\cite{mottTransitionMetallicState1961, keldyshPossibleInstabilitySemimetallic1965, Jerome1967, halperinExcitonicStateSemiconductorSemimetal1968}.
For semiconductors with a direct gap,
this phase appears when the binding energy of excitons
exceeds the band gap, as the spontaneous generation and
condensation of excitons leads to a lowering of the system's
energy. 
On the other hand, in semimetals, where
the Fermi surfaces formed by electrons and holes match each other, the attractive Coulomb interaction binds the free
electrons and holes at the Fermi level leading to BCS-type paring instability of the Fermi surfaces (FS).
Such a BCS-like condensation in momentum
space opens an energy gap at the FS, resulting in an excitonic insulating state.

The short lifetime of excitons in semiconductors often impedes the realization of exciton condensation. 
To overcome this challenge, quantum wells or van der Waals heterostructures are often used\cite{zhuExcitonCondensateSemiconductor1995, butovExcitonCondensationCoupled2003,eisensteinBoseEinsteinCondensation2004} where electrons and holes are separated in different layers. 
The real space separation of electrons and holes in these systems results in a longer lifetime for excitons, and the weak screening in 2D makes it easier for the binding to occur.
So far, a large body of experimental\cite{butovCondensationIndirectExcitons1994, foglerHightemperatureSuperfluidityIndirect2014a, duEvidenceTopologicalExcitonic2017, liExcitonicSuperfluidPhase2017, wangEvidenceHightemperatureExciton2019a, maStronglyCorrelatedExcitonic2021} and theoretical\cite{shimSpinorbitInteractionsBilayer2009, wuTheoryTwodimensionalSpatially2015, pikulinInterplayExcitonCondensation2014, xieElectricalReservoirsBilayer2018, zhuGateTuningExciton2019} studies have been conducted to investigate the properties of excitonic insulators in such 2D bilayer systems.

Although excitonic insulators have been discussed in the literature for over half a century, very few material systems have been confirmed experimentally to exhibit such exotic states. 
This is because the exciton condensation only breaks the particle-hole $U(1)$ symmetry, resulting in charge-neutral superfluidity, which is very hard to detect directly. 
In this study, we propose that the excitonic insulator in BCS limit may possess a unique breakdown mechanism, which can serve as a critical ``smoking gun'' type of experimental evidence, helping to distinguish an excitonic insulator from ordinary narrow-gap semiconductors.

\begin{figure}[!hbpt]
  \centering
  \def\svgwidth{\linewidth}
  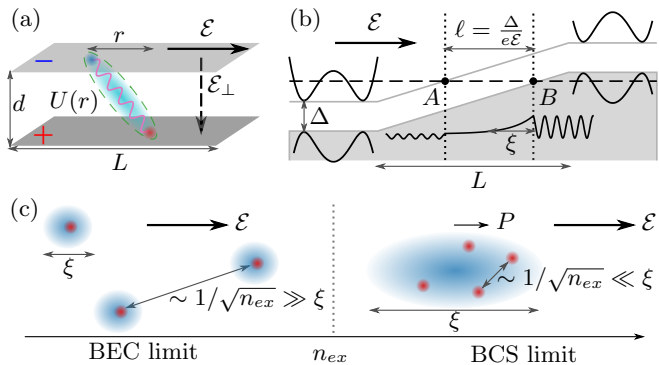
  \caption{(a) Setup of the bilayer system. 
  A vertical displacement field $\mathcal{E}_{\perp}$ is applied to equally charge the two layers with electrons and holes. 
  At the CNP, attractive interaction $U(r)$ between electrons and holes will drive the system into an excitonic insulator phase. 
  When an in-plane electric field $\mathcal{E}$ is applied, the insulating system will be polarized and even broken down.
  (b) Under rigid band assumption, inter-band Zener tunneling happens only when the in-plane bias voltage exceeds the band gap, i.e. $e\mathcal{E}L> \Delta$.
  For any energy-allowed tunneling process, there exists a classically forbidden region (from $B$ to $A$) with width $\ell=\Delta/e\mathcal{E}$ where the wavefunction decays.
  The correlation length of the gap $\xi$ characterizes the penetration depth of the wavefunction into the classically forbidden region.
  (c) The rigid band assumption is valid only when the polarization effect is negligible.
  This is true for the BEC limit where the coherence length $\xi$ is much smaller than the distance between exciton (roughly $1/\sqrt{n_{ex}}$ in 2D) and the inter-band Zener tunneling could also be viewed as single particle tunneling of each electron-hole pair.
  However, in the BCS limit, a macroscopic polarization $P$ will be induced by an external electrical field and the competition between the polarization and the condensation energies will lead to the instability of the many-body ground state, resulting in the many-body breakdown.}
  \label{fig:physical_pictures}
\end{figure}

We treat the breakdown problem of excitonic insulators by a simplified theoretical model, which contains a 2D bilayer system with a non-zero inter-layer distance, as shown in Fig. \ref{fig:physical_pictures}(a).
The application of a vertical displacement field $\mathcal{E}_{\perp}$ will result in the charging of the two layers by electrons and holes. 
If the interaction is absent, the charged bilayer would be expected to exhibit metallic behavior. 
However, the presence of an attractive interaction $U(r)$ between electrons and holes will drive the system into an excitonic insulator state at the charge neutral point (CNP). 
When an in-plane electric field $\mathcal{E}$ is gradually added, the excitonic insulator is expected to be polarized and eventually broken down.

The most well-known intrinsic breaking down mechanism for band insulators is attributed to inter-band Zener tunneling\cite{zenerTheoryElectricalBreakdown1934, esakiNewPhenomenonNarrow1958, wannierWaveFunctionsEffective1960, kaneZenerTunnelingSemiconductors1960, kaneInterbandTunneling1969}.
In an infinite system, the total energy becomes unbounded below when a uniform electric field is applied, resulting in the absence of a ground state. 
However, a finite system can still maintain an insulating stationary state at low electric fields \cite{nenciuDynamicsBandElectrons1991, souzaFirstPrinciplesApproachInsulators2002}. 
If we take the rigid band assumption and only include the electric field by a positional dependent chemical potential, the single particle Zener tunneling process can occur when the in-plane bias voltage $e\mathcal{E}L$ becomes comparable to the band gap $\Delta$ as shown in Fig. \ref{fig:physical_pictures}(b). 
This means the Zener critical field is inversely proportional to the system size $L$.
To go beyond the rigid band picture, \citet{souzaFirstPrinciplesApproachInsulators2002} consider the polarization of the occupied bands and they find the $1/L$ behavior of the Zener field still stands.

We would emphasize that this field denotes the onset of Zener tunneling when a current proportional to the tunneling probability starts to flow.
Under WKB approximation, the tunneling probability could be expressed as $\re^{-\ell/\xi}$\cite{sugimotoFieldinducedMetalinsulatorTransition2008}, where $\xi$ is the correlation length determined by the gap $\Delta$ and the tunneling length $\ell=\Delta/e\mathcal{E}$ is the width of the classically forbidden region for the Zener tunneling process.
For an excitonic insulator in the BCS limit, $\xi\approx 4v_F/\pi \Delta$ is just the coherence length of the exciton condensate (details could be found in Appendix \ref{app:zener_tunneling}).
At fixed voltage, the current is exponentially small as the system size increases\cite{seabaughLowVoltageTunnelTransistors2010, maInterbandTunnelingTwodimensional2013} due to its tunneling nature, and this current-carrying state is indeed a long-lived resonance state as discussed by \citet{souzaDynamicsBerryphasePolarization2004}.

In the BEC limit, the Zener tunneling current experiences a sharp increase when the electric field reaches $\Delta /e\xi$. 
This phenomenon can be explained as the single particle tunneling of each electron-hole pair in the center of mass system from a bound state to an extended state, similar to what happens in a normal insulator.

However, in the BCS limit, the insulating behavior in an excitonic insulator is caused by the BCS type pairing wave function, which spontaneously breaks the particle-hole $U(1)$ symmetry and gives rise to a breakdown mechanism that is specific to BCS type excitonic insulators. 
This mechanism arises from the instability of the many-body ground state caused by the competition between polarization and condensation energies. 

In the present paper, we will show that the critical field strength for this many-body breakdown is much smaller than that of the Zener breakdown (estimated roughly by $\Delta /e\xi$) in the BCS limit. 
Therefore, this unique electric breakdown feature can be considered as an important experimental signal for excitonic insulators, serving as a ``smoking gun'' to identify their presence.

\section{Polarized Mean Field Theory}

The actual breakdown scenario in excitonic insulators is complex since these two mechanisms could take effect at the same time.
To better understand the breakdown of excitonic insulators, we will utilize a self-consistent mean field theory to analyze the interplay between Zener tunneling and the many-body breakdown.

For simplicity, we will limit our analysis to the lowest conduction band in the electron layer and the highest valence band in the hole layer which are described by the electron creation operators $c^{\dagger}_{e\bm{k}}$ and $c^{\dagger}_{h\bm{k}}$, respectively.
Within the $k\cdot p$ approximation, the many-body Hamiltonian can be written as 
\begin{align}
  H&=\sum_{s=eh\bm{k}}h^0_{ss'\bm{k}}c^{\dagger}_{s\bm{k}}c_{s'\bm{k}}+e\bm{\mathcal{E}}\cdot\int\rd\bm{r}\;\Psi^{\dagger}(\bm{r})\bm{r}\Psi(\bm{r})\nonumber\\
  &+\frac{1}{2\mathcal{V}}\sum_{ss'=eh\bm{k}_1\bm{k}_2\bm{q}}V_{ss'}(\bm{q})c^{\dagger}_{s\bm{k}_1}c^{\dagger}_{s'\bm{k}_2}c_{s'\bm{k}_2+\bm{q}}c_{s\bm{k}_1-\bm{q}},\label{eq:manybody_hamiltonian}
\end{align}
where $\mathcal{V}$ is the area of the 2D system and $\Psi(\bm{r})$ is field operator defined as $\Psi(\bm{r})=\mathcal{V}^{-1/2}\sum_{s\bm{k}}\re^{i\bm{k}\cdot\bm{r}}u_{s}(\bm{r})c_{s\bm{k}}$.

The single particle Hamiltonian $h^0_{\bm{k}}$ is taken as 
\begin{equation}
  h^0_{\bm{k}}=\begin{bmatrix}
    k^2/2m_e+E_{g}-\mu_{ex} & 0\\
    0 & -k^2/2m_h 
  \end{bmatrix}.\label{eq:single_Hamiltonian}
\end{equation}
where $m_{e,h}$ are effective masses of these two kind of electrons and $E_{g}>0$ is the original band gap. 
The exciton chemical potential $\mu_{ex}$ is tuned by the vertical displacement field $\mathcal{E}_{\perp}$.
Vanishing of the off-diagonal term in Eq. \eqref{eq:single_Hamiltonian} means direct inter-layer hopping is forbidden. 
This assumption is made because we are concerned with the breakdown of an excitonic gap.
In real materials, this could be realized by symmetry constraints or just a large separation between layers.
The inter- and intra-layer interaction are taken as the Coulomb ones: $V(r)\equiv V_{s=s'}=e^2/\epsilon r$ and $U(r)\equiv V_{s\ne s'}=e^2/\epsilon\sqrt{r^2+d^2}$ whose Fourier transformations are $V(q)=2\pi e^2/\epsilon q$, $U(q)=V(q)\re^{-qd}$.

Although an in-plane field breaks translation symmetry, to describe an insulating ground state, we can always take a trial state that keeps translation symmetry as long as the field is adiabatically added (the proof is in Appendix \ref{app:trial_state}).
A trial HF state with translation symmetry at the CNP is $\ket{G}=\prod_{\bm{k}}c^{\dagger}_{v\bm{k}}\ket{\mathrm{vac}.}$, where the valence band $c^{\dagger}_{v\bm{k}}=\alpha_{\bm{k}}c^{\dagger}_{e\bm{k}}+\beta_{\bm{k}}c^{\dagger}_{h\bm{k}}$ is a linear combination of the electron and hole band with constraints $|\alpha|^2+|\beta|^2=1$. 
By using Dirac notation $\ket{v\bm{k}}=[\alpha_{\bm{k}},\beta_{\bm{k}}]^T$, energy per area becomes a functional of $\ket{v\bm{k}}$, i.e.
\begin{align}
  &\varepsilon_{tot}[\ket{v\bm{k}};\mathcal{E}]
  \equiv\frac{1}{\mathcal{V}}\bra{G}H\ket{G}\nonumber\\
  =&\frac{1}{\mathcal{V}}\sum_{s\bm{k}}h^0_{ss\bm{k}}\rho_{ss\bm{k}}
  +\frac{-e\mathcal{\mathcal{E}}}{\mathcal{V}\Delta k_{\parallel}}\mathrm{Im} \sum_{\bm{k}}\log\inp{v\bm{k}}{v\bm{k}+\Delta \bm{k}_{\parallel}}\nonumber\\
  &+\frac{2\pi e^2 n_{ex}^2d}{\epsilon}-\frac{1}{2\mathcal{V}^2}\sum_{ss'\bm{k}_1\bm{k}_2}V_{ss'}(\bm{k}_1-\bm{k}_2)\tilde{\rho}_{ss'\bm{k}_1}\tilde{\rho}_{s's\bm{k}_2},\label{eq:tot-energy}
\end{align}
where $\tilde{\rho}\equiv \rho-\rho^0$ is the density matrix relative to the initial uncharged state $\rho^0_{ss'}=\delta_{ss'}\delta_{sh}$ and $\rho$ is calculated as $\rho_{ss'\bm{k}}\equiv\bra{G}c^{\dagger}_{s'\bm{k}}c_{s\bm{k}}\ket{G}=(\ket{v\bm{k}}\bra{v\bm{k}})_{ss'}$. A general form of this functional could be found in Appendix \ref{app:energy_functionals}.

The four terms in Eq. \eqref{eq:tot-energy} could be viewed as kinetic, polarization, Hartree, and Fock energies separately.
The Hartree energy is just the charging energy of the two-layer capacitor with the electron/hole density (exciton density) $n_{ex}=1/\mathcal{V}\sum_{\bm{k}}\rho_{ee\bm{k}}$.
The relative density matrix $\tilde{\rho}$ is used in the Fock energy expression to avoid the double counting problem\cite {shimSpinorbitInteractionsBilayer2009}.
For numerical convenience, a periodic boundary condition is assumed, and the polarization energy is calculated with the help of the expectation value of many-body position operators defined on a ring geometry\cite{restaQuantumMechanicalPositionOperator1998}, which is just a discrete form of Berry phase \cite{zakBerryPhaseEnergy1989, vanderbiltElectricPolarizationBulk1993a,  King-Smith1993}. 
This form of polarization energy functional has already been used to calculate the electrical properties of insulators in the literature\cite{nunesBerryphaseTreatmentHomogeneous2001, souzaFirstPrinciplesApproachInsulators2002, iniguezFirstprinciplesStudyMathrmBiScO2003}.
On the other hand, for the open boundary problem, the polarization energy functional should be written in real space by Wannier functions\cite{nunesRealSpaceApproachCalculation1994,fernandezInitioStudyDielectric1998}.


The local minimum is found by requiring the first order derivative of $\varepsilon_{tot}$ to be zero, i.e. $\delta \varepsilon_{tot}/\delta \bra{v\bm{k}}=0$ (details are presented in Appendix \ref{app:mf_Hamiltonian}.).
This gives the mean-field Hamiltonian $h^{MF}_{\bm{k}}\equiv h^0_{\bm{k}}+h^H+h^F_{\bm{k}}+h^P_{\bm{k}}$ where
\begin{subequations}
\begin{gather}
  h^H[\ket{v\bm{k}}]=\frac{4\pi e^2 n_{ex}d}{\epsilon}(1-\rho^0),\label{eq:single_hartree}\\
  h^F_{ss'\bm{k}}[\ket{v\bm{k}}]=-\frac{1}{\mathcal{V}}\sum_{\bm{k}'}V_{s's}(\bm{k}-\bm{k}')\tilde{\rho}_{ss'\bm{k}'},\label{eq:single_fock}\\
  h^P_{\bm{k}}[\ket{v\bm{k}};\mathcal{E}]=\frac{ie\mathcal{E}}{2\Delta k_{\parallel}}\sum_{\sigma=\pm}\frac{\sigma\ket{v\bm{k}+\sigma\Delta \bm{k}_{\parallel}}\bra{v\bm{k}}}{\inp{v\bm{k}}{v\bm{k}+\sigma\Delta \bm{k}_{\parallel}}}+h.c.,\label{eq:single_pol}
\end{gather} 
\label{eq:mf_hamiltonian}
\end{subequations}
as well as the self-consistent equation
\begin{equation}
  h^{MF}_{\bm{k}}[\ket{v\bm{k}};\mathcal{E}]\ket{v\bm{k}}=\xi_{v\bm{k}}\ket{v\bm{k}}.\label{eq:self-consistent}
\end{equation}

\section{Results}

\begin{figure}[!hbpt]
  \centering
  \includegraphics[width=\linewidth]{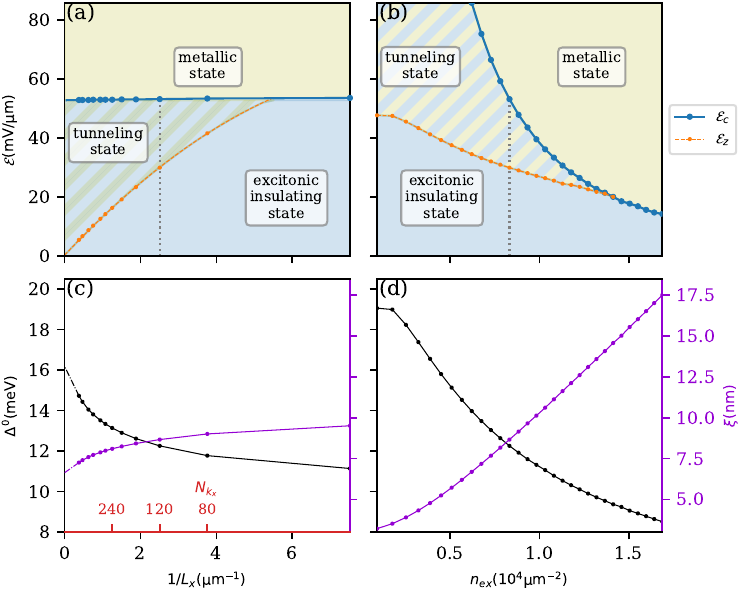}
  \caption{(a)(b) Phase diagram as a function of the in-plane electrical field $\mathcal{E}$, system size $1/L_x$ and exciton density $n_{ex}$.
  The critical field $\mathcal{E}_c$ (solid blue lines) firstly divides the entire region into a locally gapped phase and a metallic phase.
  The Zener field $\mathcal{E}_z$ (dashed orange lines) solved by $e\mathcal{E}_z L=\Delta(\mathcal{E}_z)$ marks the onset of Zener tunneling and further separates the locally gapped phase into an excitonic insulating phase and tunneling phase.
  (c)(d) Zero field band gap $\Delta^0$ (black lines, left axis) and the correlation length $\xi$ (purple lines, right axis) as functions of system size and exciton density.
  The red labels above the bottom axis of (c) mark the number of $k$ points used for the corresponding system size.
  }
  \label{fig:critical_field}
\end{figure}

In the phase diagram depicted in Fig. \ref{fig:critical_field}(a)(b), the abscissas represent the system size $1/L_{x}$ and exciton density $n_{ex}$ separately, and the vertical axis is the in-plane electric field strength $\mathcal{E}$. 
The zero-field band gap $\Delta^0$ (black line, left axis) and the correlation length $\xi$ (purple line, right axis) estimated by Eq. \eqref{eq:correlation_length} are also plotted as functions of system size and exciton density separately in Fig. \ref{fig:critical_field}(c)(d).

The parameters in our model are set as $m_{e,h}=0.3m_0$ ($m_0$ is the electron bare mass), $\epsilon=8$ and $d=100\mathrm{a.u.}\approx 5.3\mathrm{nm}$. 
The momentum space summation in Eq. \ref{eq:tot-energy} is restricted in the region $|k_{x,y}|<k_c=0.05\mathrm{a.u.}\approx 0.94\mathrm{nm}^{-1}$.
The numerical results are nearly independent of the cut-off $k_c$ when $k_c\gg k_F$ since the BCS-type condensation only occurs in a small range around $k_F$.
The size of the system is defined by the spacing of $k$-mesh as $L=2\pi/\Delta k$, so the varying of system size is realized by using different sizes of $k$-mesh.
The electrical field is applied in the $x$ direction, and the length of the system perpendicular to it is fixed at $L_y=2\pi N_{k_y}/2k_c\approx 266$nm ($N_{k_y}=80$) for numerical convenience.

In Fig. \ref{fig:critical_field}(a)(c), the exciton density is fixed at about $n_{ex}\approx 0.84\times 10^{4}\mathrm{\mu m}^{-2}$ and the number of $k$ points in the $x$ direction is taken as $N_{k_x}=40 M$ ($M$ is an integer and some used $N_{k_x}$ are marked by red texts above the bottom axis in Fig. \ref{fig:critical_field}(c)).
On the contrary, in Fig. \ref{fig:critical_field}(b)(d), the system size is fixed ($k$-mesh is fixed at $120\times 80$) and the exciton density varies.

As is shown in Fig. \ref{fig:critical_field}(c)(d), the correlation length $\xi$ is about $10$nm within the range of the parameters we consider.
The correlation length $\xi$ is much smaller than the system size $L_x$ along the direction of the electrical field, which means tunneling current at the onset of Zener tunneling $I\propto \re^{-L_x/\xi}$ is negligible.

To overcome the Zener instability of the energy functional for the electrical field in the range $\Delta/eL_x\sim \Delta/e\xi$, the polarization Hamiltonian $h^P_{\bm{k}}$ and the polarization energy are always evaluated on the coarse $40\times 80$ mesh.
For an original $40 M\times 80$ $k$-mesh, this is equivalent to dividing the system into $M$ copies with size $L_x= L_0 \approx 133\mathrm{nm}$.
Thus the Zener tunneling process whose tunneling length $\ell$ satisfies $ML_0> \ell >L_0> \xi$ is ignored.
This approximation is reasonable since the tunneling probability $\re^{-\ell/\xi}$ for such process is smaller than $\re^{-L_0/\xi}\approx 10^{-3}$.

The blue lines in Fig. \ref{fig:critical_field}(a)(b) represent the critical field $\mathcal{E}_c$ accounting for the many-body breakdown of the excitonic gap, which divides the entire region into a metallic phase and a locally gapped phase.
By solving $e\mathcal{E}_zL_x=\Delta(\mathcal{E}_z)$, the minimum field $\mathcal{E}_z$ required for Zener tunneling is obtained and plotted by the orange lines and further separates the locally gapped phase into an excitonic insulating phase and a tunneling phase.
In the excitonic insulating phase, the system is fully gapped, and no current flows.
In the tunneling phase, an exponentially small Zener tunneling current appears while the system is still locally gapped.
In the metallic phase, the excitonic gap is destroyed, the system becomes highly conductive and the resistivity-temperature ($R-T$) curve becomes typical metallic.

To understand the breakdown phase transition, let's examine the stability of the local minimum, which is realized by calculating the second-order derivatives (Hessian matrix) of the energy functional.

Assume we are in the region of insulating state, so the local minimum $\ket{v\bm{k};\mathcal{E}}$ could be found by our self-consistent procedure.
The self consistent equation at the mean field solution reads $h^{MF}_{\bm{k}}[\ket{v\bm{k};{\mathcal{E}}};{\mathcal{E}}]\ket{i\bm{k};{\mathcal{E}}}=\xi_{i\bm{k},\mathcal{E}}\ket{i\bm{k};{\mathcal{E}}}$,
where $\ket{c\bm{k};{\mathcal{E}}},\ket{v\bm{k};{\mathcal{E}}}$ are conduction and valance bands with band energies $\xi_{c\bm{k},\mathcal{E}}>\xi_{v\bm{k},\mathcal{E}}$.
At the local minimum, the trial HF state could be re-parameterized as
\begin{equation}
  \ket{v'\bm{k};\mathcal{E}}=({\ket{v\bm{k};\mathcal{E}}+f_{\bm{k}}\ket{c\bm{k};\mathcal{E}}})/{\sqrt{1+|f_{\bm{k}}|^2}}.
\end{equation}
This parametrization is complete and unconstrained ($f_{\bm{k}}$ is an arbitrary complex-valued function).
Then the total energy becomes a functional of $f_{\bm{k}}$, i.e. $\varepsilon_{tot}[f_{\bm{k}},f_{\bm{k}}^*;\mathcal{E}]\equiv \varepsilon_{tot}[\ket{v'\bm{k};\mathcal{E}};\mathcal{E}]$,
and the Hessian matrix at this point is
\begin{equation}
  \mathrm{H}_{\bm{k}\bm{k}'}=\begin{pmatrix}
    \frac{\delta \varepsilon_{tot}}{\delta \mathrm{Re}f_{\bm{k}}\delta \mathrm{Re}f_{\bm{k}'}}
      & \frac{\delta \varepsilon_{tot}}{\delta \mathrm{Re}f_{\bm{k}}\delta \mathrm{Im}f_{\bm{k}'}}\\
      \frac{\delta \varepsilon_{tot}}{\delta \mathrm{Im}f_{\bm{k}}\delta \mathrm{Re}f_{\bm{k}'}} & \frac{\delta \varepsilon_{tot}}{\delta \mathrm{Im}f_{\bm{k}}\delta \mathrm{Im}f_{\bm{k}'}}
  \end{pmatrix}\Bigg|_{f_{\bm{k}}=0}.\label{eq:Hessian}
\end{equation}
The details and specific expression of $\mathrm{H}_{\bm{k}\bm{k}'}$ could be found in Appendix \ref{app:hessian}.

If $f_{\bm{k}}$ is small, $\ket{v'\bm{k};\mathcal{E}}$ is approximated by $\ket{v'\bm{k};\mathcal{E}}\sim\ket{v\bm{k};\mathcal{E}}+f_{\bm{k}}\ket{c\bm{k};\mathcal{E}}$.
Such a form can be viewed as the low-energy excitation modes in the variational parameter space illustrated above.
By diagonalizing the Hessian matrix, the eigenmodes for the low-energy excitations $\sum_{\bm{k}'}\mathrm{H}_{\bm{k}\bm{k}'}f^{\lambda}_{\bm{k}'}=\lambda f^{\lambda}_{\bm{k}'}$
can be obtained.
For convenience, the eigenmodes in the following text are normalized by $f^{\lambda}_{\bm{k}}\to f^{\lambda}_{\bm{k}}/\sqrt{\sum_{\bm{k}}|f_{\bm{k}}^{\lambda}|^2}$.

\begin{figure}[!hbpt]
  \centering
  \includegraphics[width=\linewidth]{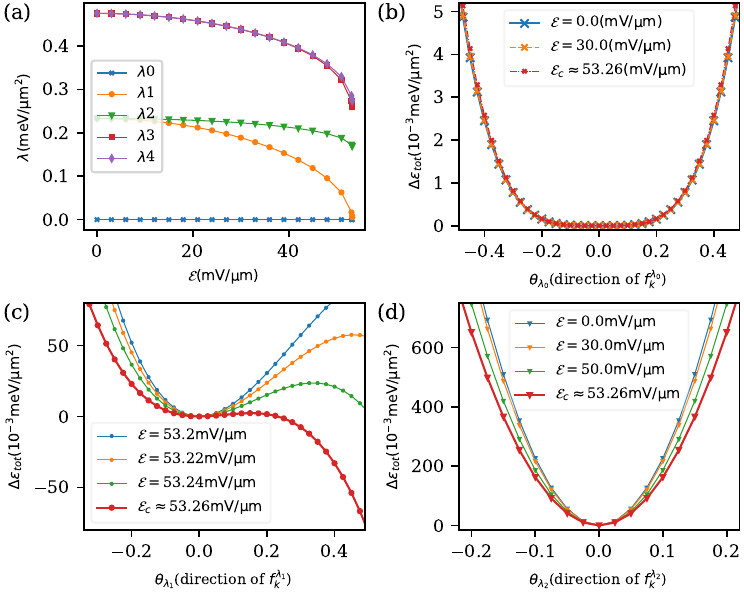}
  \caption{(a) The smallest five eigenvalues of the Hessian matrix Eq. \eqref{eq:Hessian} as a function of the electric field.
  (b)(c)(d) Total energy difference Eq. \eqref{eq:energy_diff} as a function of the electric field and excitation amplitudes along the directions $f^{\lambda_{0,1,2}}_{\bm{k}}$ in the variational parameter space. 
  The excitation amplitudes are used as the horizontal axes while different field strengths are represented by different color lines.
  These data are generated on a $120\times 80$ $k$-mesh with exciton exciton density $n_{ex}\approx 0.84\times10^{4}\mathrm{\mu m}^{-2}$ (along the dashed gray line in Fig. \ref{fig:critical_field}(a)(b)).}
  \label{fig:losing_minimum}
\end{figure}

On a $120\times 80$ $k$-mesh with exciton density $n_{ex}\approx 0.84\times10^{4}\mathrm{\mu m}^{-2}$ (dashed gray line in Fig. \ref{fig:critical_field}(a)(b)), the total energy functional is analyzed, and the results are shown in Fig. \ref{fig:losing_minimum}.
In Fig. \ref{fig:losing_minimum}(a), we plot the smallest few eigenvalues $\lambda_{0-4}$ of the Hessian matrix Eq. \eqref{eq:Hessian} as functions of field strength.
By taking trial HF state as $\ket{v'\bm{k};\mathcal{E};\theta_{\lambda_i}}\propto \ket{v\bm{k};\mathcal{E}}+\theta_{\lambda_i} f^{\lambda_i}_{\bm{k}}\ket{c\bm{k};\mathcal{E}}$,
the total energy difference between the trial state and the HF ground state along the directions $f^{\lambda_i}_{\bm{k}}$ in the variational parameter space is evaluated as
\begin{equation}
  \Delta \varepsilon_{tot}(\mathcal{E},\theta_{\lambda_i})\equiv \varepsilon_{tot}[\ket{v'\bm{k};\mathcal{E};\theta_{\lambda_i}}]-\varepsilon_{tot}[\ket{v\bm{k};\mathcal{E}}]\label{eq:energy_diff}.
\end{equation}
Using the lowest three eigenmodes $f_{\bm{k}}^{\lambda_{0,1,2}}$ for example, the total energy difference as a function of the electric field $\mathcal{E}$ and excitation amplitudes $\theta_{\lambda_i}$ is plotted in Fig. \ref{fig:losing_minimum}(b)(c)(d).
In these plots, the horizontal axes are the amplitudes of those eigenmodes, while different electric field strengths are represented by different color lines.

There is a consistent zero mode $\lambda_0$ for any electric field strength, as shown in Fig. \ref{fig:losing_minimum}(a).
However, the behaviors of the total energy functional along the direction $f^{\lambda_0}_{\bm{k}}$ in Fig. \ref{fig:losing_minimum}(b) indicates that it's not a ``breaking down mode'' because the high-order derivatives of the total energy functional along this direction are always positive. 
Such a zero mode is exactly the Goldstone mode related to phase fluctuation of the exciton condensate and accounts for the exciton superfluidity (see details in Appendix \ref{app:Goldstone_mode}).

The real breaking down direction in parameter space is $f_{\bm{k}}^{\lambda_1}$ as shown in Fig. \ref{fig:losing_minimum}(c).
When the electric field is small, all eigenvalues of the Hessian matrix(except the Goldstone mode $\lambda_0$) satisfy $\lambda>\lambda_1>0$, which means the solution is indeed a local minimum.
As the electric field approaches the critical field strength $\mathcal{E}_{c}$, the eigenvalue of the breakdown mode $\lambda_1$ approaches $0$ and the excitonic insulator ground state becomes unstable as the local minimum turns into a saddle point. 
Such a many-body breakdown mechanism is completely different from traditional Zener tunneling and the corresponding critical field strength can be much weaker than the one for Zener tunneling as discussed in the following section.

\section{Discussion}

\begin{figure}[!hbpt]
  \centering
  \includegraphics[width=\linewidth]{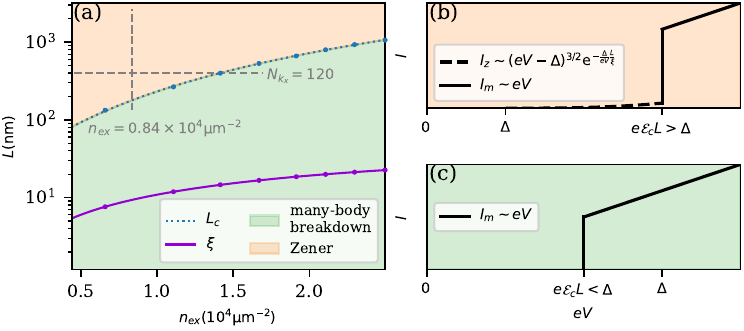}
  \caption{(a) The critical length $L_c$ where the Zener field $\mathcal{E}_z$ equals the critical field $\mathcal{E}_c$ for the many-body breakdown in Fig. \ref{fig:critical_field}(a) is plotted as a function of exciton density by the dotted blue line, which separates the $n_{ex}-L$ plane into two regions, i.e. the many-body breakdown region (green color) and the Zener region (orange color).
  The correlation length of the excitonic gap $\xi$ given by Eq. \eqref{eq:correlation_length} is also plotted by the purple line for reference.
  The two dashed gray lines mark the paths along which Fig. \ref{fig:critical_field} is generated.
  (b)(c) $I-V$ characteristics for the excitonic insulator in the many-body breakdown region and Zener region separately.
  }
  \label{fig:IV_curve}
\end{figure}

The results in Fig. \ref{fig:critical_field}(a)(b) indicate that the critical field $\mathcal{E}_c$ for the many-body breakdown is nearly independent of the system size and decreases dramatically with the increase in exciton density.
This is reasonable since with the increase in exciton density, the binding between electron and hole becomes weaker and the excitonic insulator will eventually turn into a quantum electron-hole plasma state\cite{mottTransitionMetallicState1961,nikolaevTheoryExcitonicMott2008,asanoExcitonMottPhysics2014}.
Then the intersection point of the two critical fields $\mathcal{E}_c$ and $\mathcal{E}_z$ (intersection points of the blue and orange lines in Fig. \ref{fig:critical_field}(a)(b)) gives a critical length roughly estimated by
\begin{equation}
  L_c\sim \Delta/e\mathcal{E}_c,\label{eq:critical_line}
\end{equation}
which separates the $n_{ex}-L$ plane into two regions as illustrated in Fig. \ref{fig:IV_curve}(a).
And the $I-V$ characteristic may behave differently in the two regions.

In the many-body breakdown region below the line of critical length $L_c$ (the green color region in Fig. \ref{fig:IV_curve}(a)), the excitonic gap is disrupted before the onset of inter-band Zener tunneling.
As the electrical field increases, the BCS-type exciton condensation wave function will lose stability and exhibit a typical first-order transition feature. 
After this transition, the system becomes gapless and highly conductive, and a metallic current $I_{m}\propto eV$ will flow in the system.

In the Zener region where $L\gg L_c$ and $\tilde{\mathcal{E}}_z\ll \tilde{\mathcal{E}}_{ex}$ (the orange color region in Fig. \ref{fig:IV_curve}(a)), a tunneling current will first appear when the in-plane bias voltage exceeds the band gap.
Fot gate voltage in the range $\Delta\sim e\mathcal{E}_{z}L \ll eV<e\mathcal{E}_{c}L$, this current is in the form of
\begin{equation}
  I_{z}(eV\equiv e\mathcal{E}L)\sim (eV-\Delta)^{3/2}\re^{-\frac{\Delta}{eV}\frac{L}{\xi}}.
\end{equation}
The exponential factor $\re^{-\Delta L/eV\mathcal{\xi}}$ is the WKB tunneling probability and power term $(eV-\Delta)^{3/2}$ arises from the density of states of the tunneling channels in 2D systems (details could be found in Appendix \ref{app:zener_tunneling}).
Different from the metallic current, the tunneling current exhibits a different $R-T$ characteristic, i.e. the current increases when the temperature rises.

The tunneling current persists until the field strength reaches the critical field of many-body breakdown, after which the excitonic gap disappears and a metallic current $I_m\propto eV$ appears replacing the Zener tunneling current $I_z$.
However, even at the critical field $\mathcal{E}_{c}$, the tunneling current $I_z(\mathcal{E}=\mathcal{E}_c) \propto \re^{-L_c/\xi}$ in the BCS limit is still exponentially small as the critical length $L_c$ is nearly two orders larger than correlation length $\xi$ as is shown in Fig. \ref{fig:IV_curve}(a).
This means a switching phenomenon of the $I-V$ curve is still observable even in the Zener region.

From the discussion above, the typical $I-V$ characteristics in the two regions are schematically illustrated in Fig. \ref{fig:IV_curve}(b)(c).

We notice that \citet{sugimotoFieldinducedMetalinsulatorTransition2008} also proposed a breaking down mechanism in correlated insulators which has a threshold field much smaller than that for Zener breakdown.
However, the mechanism in their work is distinct from the many-body breakdown mechanism proposed in our work.
The many-body breakdown is intrinsic for an excitonic insulator while the critical field in their work is related to the extrinsic relaxation time.
Besides, the typical $I-V$ curve for an excitonic insulator as illustrated in Fig. \ref{fig:IV_curve}(b)(c) has size dependence which is already observed by the experiments of \citet{yangUnconventionalCorrelatedInsulator2023}.

At last, the many-body breakdown mechanism is a breakdown of the electronic band structure and has nearly no influence on the lattice, which means the breaking-down process is reversible and the switching phenomenon of the $I-V$ characteristic is promising for practical usage.

\begin{acknowledgments}

We thank Prof. Zheng Vitto Han, Naoto Nagaosa, and Wan Yao for their helpful
discussions. 
X.D. acknowledges financial support from the
Hong Kong Research Grants Council (Project No. 16309020).

\end{acknowledgments}

\appendix

\section{The Trial State}\label{app:trial_state}
We first prove that under a uniform electric field, the many-body state will keep its lattice translation symmetry at all times.

A many-body state $\ket{\mathit{\Psi};t}$ is said to have lattice translation symmetry if and only if the wavefunction satisfies
\begin{equation}
  \mathit{\Psi}(\bm{r}_1,+\bm{R}_0,\cdots,\bm{r}_{N_{e}}+\bm{R}_0;t)\nonumber
  =\re^{i\phi}\mathit{\Psi}(\bm{r}_1,\cdots,\bm{r}_{N_{e}};t),\label{eq:lattice-translation-symmtry}
\end{equation}
where $N_{e}$ is the total number of electrons and $\bm{R}_0$ is arbitrary lattice vector.

The many-body Schr\"odinger equation in length gauge (using a scalar field $\varphi=e\bm{\mathcal{E}}\cdot \bm{r}$ to include electric field) is written as
\begin{align}
  &i\partial_{t}\mathit{\Psi}^{E}(\bm{r}_1,\cdots,\bm{r}_{N_{e}};t)
  =\Big\{\sum_{i=1}^{N_{e}}\left[h^0(-i\nabla_{\bm{r}_i},\bm{r}_i)+e\bm{\mathcal{E}}\cdot \bm{r}_i\right]\nonumber\\
  &\quad +\sum_{1\le i<j\le N_{e}}V(\bm{r}_i-\bm{r}_j)\Big\}\mathit{\Psi}^{E}(\bm{r}_1,\cdots,\bm{r}_{N_{e}};t),
\end{align}
which seems to break lattice translation symmetry.
However, by taking gauge transformation of the electric field $\partial_t\bm{A}(t)=-\bm{\mathcal{E}}$ and defining
\begin{equation}
  \mathit{\Psi}^{A}(\bm{r}_1,\cdots,\bm{r}_{N_{e}};t)=\re^{-i\sum_{i=1}^{N_{e}}e\bm{A}(t)\cdot\bm{r}_i}\mathit{\Psi}^{E}(\bm{r}_1,\cdots,\bm{r}_{N_{e}};t),
\end{equation}
we find that the Schr\"odinger equation for $\ket{\mathit{\Psi}^{A}}$ becomes
\begin{align}
  &i\partial_{t}\mathit{\Psi}^{A}(\bm{r}_1,\cdots,\bm{r}_{N_{e}};t)
  =\Big[\sum_{i=1}^{N_{e}}h^0(-i\nabla_{\bm{r}_i}+e\bm{A}(t),\bm{r}_i)\nonumber\\
  &\quad +\sum_{1\le i<j\le N_{e}}V(\bm{r}_i-\bm{r}_j)\Big]\mathit{\Psi}^{A}(\bm{r}_1,\cdots,\bm{r}_{N_{e}};t),\label{eq:many-body-eq-velocity-gauge}
\end{align}
which keeps the lattice translation symmetry.
So starting from a many-body state $\ket{\mathit{\Psi}^0}$ with lattice translation symmetry, the many-body state $\ket{\mathit{\Psi}^{A};t}$ as well as $\ket{\mathit{\Psi}^{E};t}$ will have lattice translation symmetry at any time:
\begin{align}
  &\mathit{\Psi}^{E}(\bm{r}_1+\bm{R}_0,\cdots,\bm{r}_{N_{e}}+\bm{R}_0;t)\nonumber\\
  =&\re^{i\sum_{i=1}^{N_{e}}e\bm{A}(t)\cdot(\bm{r}_i+\bm{R}_0)}\mathit{\Psi}^{A}(\bm{r}_1+\bm{R}_0,\cdots,\bm{r}_{N_{e}}+\bm{R}_0;t)\nonumber\\
  =&\re^{iN_{e}e\bm{A}(t)\cdot\bm{R}_0+i\phi_A}\re^{i\sum_{i=1}^{N_{e}}\bm{A}(t)\cdot\bm{r}_i}\mathit{\Psi}^{A}(\bm{r}_1,\cdots,\bm{r}_{N_{e}};t)\nonumber\\
  =&\re^{iN_{e}e\bm{A}(t)\cdot\bm{R}_0+i\phi_A}\mathit{\Psi}^{E}(\bm{r}_1,\cdots,\bm{r}_{N_{e}};t).
\end{align}

When treating a static uniform electric field, as long as the field is adiabatically turned on, a trial HF state with lattice translation symmetry could be safely assumed.
For insulators, this state is written as
\begin{equation}
  \ket{GS}=\prod_{n=1}^{n_{e}}\prod_{\bm{k}\in BZ}c^{\dagger}_{n\bm{k}}\ket{\mathrm{vac}.},
\end{equation}
where $n_{e}$ is electron per cell and $\ket{\mathrm{vac}.}$ is vacuum state.
$c^{\dagger}_{n\bm{k}}$ is creation operators of Bloch electron with wavefunction
\begin{equation}
  \psi_{n\bm{k}}(\bm{r})=\{\Psi(\bm{r}),c^{\dagger}_{n\bm{k}}\}=\frac{1}{\sqrt{\mathcal{V}}}\re^{i\bm{k}\cdot\bm{r}}u_{n\bm{k}}(\bm{r}),
\end{equation}
where $\mathcal{V}=\mathcal{N}v_{c}$ is the volume of the system, $\mathcal{N}$ is the number of unit cells and $v_{c}$ is cell volume.

As electron creation operators, $c^{\dagger}_{n\bm{k}}$ should satisfy 
\begin{equation}
  \{c^{\dagger}_{m\bm{k}},c_{n\bm{k}'}\}=\delta_{mn}\delta_{\bm{k}\bm{k}'},
\end{equation}
which forces the corresponding Bloch functions to be orthonormal, i.e.
\begin{subequations}
  \begin{gather}
    \inp{\psi_{m\bm{k}}}{\psi_{n\bm{k}'}}=\int\rd\bm{r}\;\psi^*_{m\bm{k}}(\bm{r})\psi_{n\bm{k}'}(\bm{r})=\delta_{mn}\delta_{\bm{k}\bm{k}'},\\
    \inp{u_{m\bm{k}}}{u_{n\bm{k}}}=\frac{1}{v_{c}}\int_{cell}\rd\bm{r}\;u^*_{m\bm{k}}(\bm{r})u_{n\bm{k}}(\bm{r})=\delta_{mn}.
  \end{gather}
\end{subequations}

\section{Polarized HF Energy Functional}\label{app:energy_functionals}
In this part, a general form of the polarized HF energy as a functional of occupied Bloch states will be derived.

Using field operator $\Psi(\bm{r})$, the second quantization form of the single particle (kinetic and potential energy), polarization, and interaction Hamiltonians are written as
\begin{gather}
  H_{0}=\int\rd\bm{r}\;\Psi^{\dagger}(\bm{r})h^0(-i\nabla_{\bm{r}},\bm{r})\Psi(\bm{r}),\\
  H_{P}=e\bm{\mathcal{E}}\cdot\int\rd\bm{r}\;\Psi^{\dagger}(\bm{r})\bm{r} \Psi(\bm{r}),\\
  H_{I}=\frac{1}{2}\int\rd\bm{r}_1\rd\bm{r}_2\;\Psi^{\dagger}(\bm{r}_1)\Psi^{\dagger}({\bm{r}_2})V(\bm{r}_1-\bm{r}_2)\Psi(\bm{r}_2)\Psi(\bm{r}_1).
\end{gather}

Matrix elements of the single-particle density operator $\hat{\rho}$ under position basis are calculated as
\begin{align}
  \rho(\bm{r},\bm{r}')=&\bra{GS}\Psi^{\dagger}(\bm{r}')\Psi({\bm{r}})\ket{GS}\nonumber\\
  =&\sum_{n=1}^{n_{e}}\sum_{\bm{k}\in BZ}\{c_{n\bm{k}},\Psi^{\dagger}(\bm{r}')\}\{\Psi(\bm{r}),c^{\dagger}_{n\bm{k}}\}\nonumber\\
  =&\sum_{n=1}^{n_{e}}\sum_{\bm{k}\in BZ}\psi_{n\bm{k}}(\bm{r})\psi^*_{n\bm{k}}(\bm{r}').
\end{align}
Then its $k$-dependent counterpart is defined by
\begin{equation}
  \hat{\rho}_{\bm{k}}=\mathcal{N}\re^{-i\bm{k}\cdot\hat{\bm{r}}}\hat{\rho}\re^{i\bm{k}\cdot\hat{\bm{r}}}=\sum_{i=1}^{n_{e}}\ket{u_{n\bm{k}}}\bra{u_{n\bm{k}}}.\label{eq:density-operator}
\end{equation}
Need to notice that the single particle Hilbert space $\mathcal{H}$ of $\hat{\rho}$ is all kinds of functions while the Hilbert space $\mathcal{H}_{\bm{k}}$ of $\hat{\rho}_{\bm{k}}$ is only the cell-periodic functions.
That's why the prefactor $\mathcal{N}$, number of cells, appears in the definition of $\hat{\rho}_{\bm{k}}$ in Eq. \eqref{eq:density-operator}.
And we will see the single-particle and interaction energies could be expressed as functionals of $\hat{\rho}_{\bm{k}}$ and therefore functionals of occupied states $\ket{u_{n\bm{k}}}$.

The single-particle part is
\begin{align}
  E_0\equiv&\bra{GS}H_{0}\ket{GS}\nonumber\\
  =&\int\rd\bm{r}\rd\bm{r}'\;\delta(\bm{r}-\bm{r}')h^0(-i\nabla_{\bm{r}},\bm{r})\rho(\bm{r},\bm{r}')\nonumber\\
  =&\sum_{n=1}^{n_{e}}\sum_{\bm{k}\in BZ}\int\rd\bm{r}\;\psi^{*}_{n\bm{k}}(\bm{r})h^0(-i\nabla_{\bm{r}},\bm{r})\psi_{n\bm{k}}(\bm{r})\nonumber\\
  =&\sum_{\bm{k}\in BZ}\mathrm{Tr}[\hat{h}^0_{\bm{k}}\hat{\rho}_{\bm{k}}],\label{eq:single-energy}
\end{align}
where $\hat{h}^0_{\bm{k}}=\re^{-i\bm{k}\cdot{\bm{r}}}\hat{h}^0(\hat{\bm{p}},\hat{\bm{r}})\re^{i\bm{k}\cdot\hat{\bm{r}}}=\hat{h}^0(\hat{\bm{p}}+\bm{k},\hat{\bm{r}})$ is the $k$-dependent single particle Hamiltonian acting on cell-periodic functions with matrix elements
\begin{equation}
  h^{0}_{mn\bm{k}}\equiv\frac{1}{v_{c}}\int_{cell}\rd\bm{r}\; u^{*}_{m\bm{k}}(\bm{k})h^0(-i\nabla_{\bm{r}}+\bm{k},\bm{r})u_{n\bm{k}}(\bm{r}).
\end{equation}

Similarly, the interaction part is evaluated with the help of Wick's theorem
\begin{align}
  &\bra{GS}H_{I}\ket{GS}\nonumber\\
  =&\frac{1}{2}\int\rd\bm{r}\rd\bm{r}'\;V(\bm{r}-\bm{r}')\ave{\Psi^{\dagger}(\bm{r})\Psi^{\dagger}(\bm{r}')\Psi(\bm{r}')\Psi(\bm{r})}\nonumber\\
  =&\frac{1}{2}\int\rd\bm{r}\rd\bm{r}'\;V(\bm{r}-\bm{r}')[\rho(\bm{r},\bm{r})\rho(\bm{r}',\bm{r}')-\rho(\bm{r}',\bm{r})\rho(\bm{r},\bm{r}')]\nonumber\\
  =&\frac{1}{2\mathcal{V}}\sum_{\bm{q}}V(\bm{q})\int\rd\bm{r}\rd\bm{r}'\;\re^{i\bm{q}\cdot(\bm{r}-\bm{r}')}\rho(\bm{r},\bm{r})\rho(\bm{r}',\bm{r}')\nonumber\\
  &-\frac{1}{2\mathcal{V}}\sum_{\bm{q}}V(\bm{q})\int\rd\bm{r}\rd\bm{r}'\;\re^{i\bm{q}\cdot(\bm{r}-\bm{r}')}\rho(\bm{r}',\bm{r})\rho(\bm{r},\bm{r}'),\label{eq:HF}
\end{align}
where $V(\bm{q})\equiv\int\rd\bm{r}\; V(\bm{r})\re^{-i\bm{q}\cdot\bm{r}}$ is the Fourier transformation of $V(\bm{r})$.
The first part in Eq. \eqref{eq:HF} is the Hartree energy and is simplified as 
\begin{align}
  E_H=&\frac{1}{2\mathcal{V}}\sum_{\bm{q}}V(\bm{q})\int\rd\bm{r}\rd\bm{r}'\;\re^{i\bm{q}\cdot(\bm{r}-\bm{r}')}\rho(\bm{r},\bm{r})\rho(\bm{r}'\bm{r}')\nonumber\\
  =&\frac{1}{2\mathcal{V}}\sum_{\bm{q}}V(\bm{q})\int\rd\bm{r}\;\re^{i\bm{q}\cdot\bm{r}}\sum_{n=1}^{n_{e}}\sum_{\bm{k}_1\in BZ}\psi^*_{n\bm{k}_1}(\bm{r})\psi_{n\bm{k}_1}(\bm{r})\nonumber\\
  &\times\int\rd\bm{r}'\;\re^{-i\bm{q}\cdot\bm{r}'}\sum_{m=1}^{n_{e}}\sum_{\bm{k}_2\in BZ}\psi^*_{m\bm{k}_2}(\bm{r}')\psi_{m\bm{k}_2}(\bm{r}')\nonumber\\
  =&\frac{1}{2\mathcal{V}}\sum_{\bm{k}_i\in BZ}\sum_{\bm{q}}V(\bm{q})\delta_{\bm{q}\bm{G}}\mathrm{Tr}[\re^{i\bm{q}\cdot\hat{\bm{r}}}\hat{\rho}_{\bm{k}_1}]\mathrm{Tr}[\re^{-i\bm{q}\cdot\hat{\bm{r}}}\hat{\rho}_{\bm{k}_2}]\nonumber\\
  =&\frac{1}{2\mathcal{V}}\sum_{\bm{k}_i\in BZ,\bm{G}}V(\bm{G})\mathrm{Tr}[\re^{i\bm{G}\cdot\hat{\bm{r}}}\hat{\rho}_{\bm{k}_1}]\mathrm{Tr}[\re^{-i\bm{G}\cdot\hat{\bm{r}}}\hat{\rho}_{\bm{k}_2}].\label{eq:hartree-energy}
\end{align}
where $\bm{G}$ is reciprocal vector.
The $\re^{i\bm{G}\cdot\hat{\bm{r}}}$ term in Eq. \eqref{eq:hartree-energy} should be understood as a single particle operator that acts on $\ket{u_{n\bm{k}}}$ as 
\begin{equation}
  \bra{\bm{r}}\re^{i\bm{G}\cdot\hat{\bm{r}}}\ket{u_{n\bm{k}}}=\re^{i\bm{G}\cdot\bm{r}}u_{n\bm{k}}(\bm{r})=u_{n\bm{k}-\bm{G}}(\bm{r})=\inp{\bm{r}}{u_{n\bm{k}-\bm{G}}}.
\end{equation}
The second part in Eq. \eqref{eq:HF} is the Fock energy 
\begin{align}
  E_{F}=&-\frac{1}{2\mathcal{V}}\sum_{\bm{q}}V(\bm{q})\int\rd\bm{r}\rd\bm{r}'\;\re^{i\bm{q}\cdot(\bm{r}-\bm{r}')}\nonumber\\
  &\times\sum_{m,n=1}^{n_{e}}\sum_{\bm{k}_i\in BZ}\psi_{n\bm{k}_1}(\bm{r}')\psi^*_{n\bm{k}_1}(\bm{r})\psi_{m\bm{k}_2}(\bm{r})\psi^*_{m\bm{k}_2}(\bm{r}')\nonumber\\
  =&\frac{-1}{2\mathcal{V}}\sum_{\bm{k}_i\in BZ,\bm{q}}V(\bm{q})\delta_{\bm{q}\bm{k}_1-\bm{k}_2+\bm{G}}\mathrm{Tr}[\re^{-i\bm{G}\cdot\hat{\bm{r}}}\hat{\rho}_{\bm{k}_1}\re^{i\bm{G}\cdot\hat{\bm{r}}}\hat{\rho}_{\bm{k}_2}]\nonumber\\
  =&\frac{-1}{2\mathcal{V}}\sum_{\bm{k}_i\in BZ,\bm{G}}V(\bm{k}_1-\bm{k}_2+\bm{G})\mathrm{Tr}[\re^{-i\bm{G}\cdot\hat{\bm{r}}}\hat{\rho}_{\bm{k}_1}\re^{i\bm{G}\cdot\hat{\bm{r}}}\hat{\rho}_{\bm{k}_2}].\label{eq:fock-energy}
\end{align}

The polarization energy can't be expressed by density operator $\hat{\rho}_{\bm{k}}$ but is still a functional of occupied states
\begin{align}
  E_{P}\equiv& \bra{GS}H_{P}\ket{GS}\nonumber\\
  =&e\bm{\mathcal{E}}\cdot\int\rd\bm{r}\;\bm{r}\rho(\bm{r},\bm{r})\nonumber\\
  =&e\bm{\mathcal{E}}\cdot\sum_{n=1}^{N_{e}}\sum_{\bm{k},\bm{k}'}\delta_{\bm{k},\bm{k}'}\int\rd\bm{r}\;\psi^*_{n\bm{k}'}(\bm{r})\bm{r}\psi_{n\bm{k}}(\bm{r})\nonumber\\
  =&e\bm{\mathcal{E}}\cdot\sum_{n=1}^{N_{e}}\sum_{\bm{k},\bm{k}'}\delta_{\bm{k}\bm{k}'}\times \left[-i\nabla_{\bm{k}}\int\rd\bm{r}\;\psi^{*}_{n\bm{k}'}(\bm{r})\psi_{n\bm{k}}(\bm{r})\right.\nonumber\\
  &\left.+\frac{1}{\mathcal{V}}\int\rd\bm{r}\;\re^{i(\bm{k}-\bm{k}')\cdot\bm{r}}u^*_{n\bm{k}'}(\bm{r})i\nabla_{\bm{k}}u_{n\bm{k}}(\bm{r})\right]\nonumber\\
  =&e\bm{\mathcal{E}}\cdot\sum_{n=1}^{N_{e}}\sum_{\bm{k},\bm{k}'}\delta_{\bm{k}\bm{k}'}\left[-i\nabla_{\bm{k}}\delta_{\bm{k}\bm{k}'}+\inp{u_{n\bm{k}}}{i\nabla_{\bm{k}}u_{n\bm{k}}}\right]\nonumber\\
  =&\sum_{n=1}^{n_{e}}\sum_{\bm{k}}\bra{u_{n\bm{k}}}ie\bm{\mathcal{E}}\cdot\nabla_{\bm{k}}\ket{u_{n\bm{k}}}.\label{eq:pol-energy-continuous}
\end{align}
This result is consistent with the Berry phase definition of polarization.
For a finite-size system with periodic boundary conditions, the polarization, and the polarization energy should be written with the discrete form of Berry phase as\cite{restaQuantumMechanicalPositionOperator1998} 
\begin{equation}
  E_P=\frac{-e\mathcal{E}}{\Delta k_{\parallel}}\mathrm{Im}\sum_{\bm{k}}\log\det S(\bm{k},\bm{k}+\Delta\bm{k}_{\parallel}),\label{eq:pol-energy-discrete}
\end{equation}
where $|\Delta \bm{k}_{\parallel}|=2\pi/L$ and is along the direction of electric field. The overlap matrix $S$ is defined as
\begin{equation}
  S_{mn}(\bm{k},\bm{k}')=\inp{u_{m\bm{k}}}{u_{n\bm{k}'}},\;m,n=1,2,\cdots,n_{e}.
\end{equation}

\section{Mean Field Hamiltonian and Self-consistent Equation}\label{app:mf_Hamiltonian}

The total energy as a functional of occupied bands $\{\ket{u_{n\bm{k}}}\}_{n=1}^{n_{e}}$ is written as
\begin{equation}
  E_{tot}[\ket{u_{n\bm{k}}};\mathcal{E}]=E_{0}[\hat{\rho}_{\bm{k}}]+E_{HF}[\hat{\rho}_{\bm{k}}]+E_{P}[\ket{u_{n\bm{k}}};\mathcal{E}]
\end{equation}
and the stationary state is found by minimize $E_{tot}$ with constraints
\begin{equation}
  \inp{u_{m\bm{k}}}{u_{n\bm{k}}}=\delta_{mn}.
\end{equation}
By introduce Lagrange multipliers $\xi_{n\bm{k}}$, the constrained minimization of $E_{tot}$ is transformed into an unconstrained minimization of 
\begin{equation}
  F[\ket{u_{n\bm{k}}};\mathcal{E}]\equiv E_{tot}[\ket{u_{n\bm{k}}};\mathcal{E}]+\sum_{n\bm{k}}\xi_{n\bm{k}}(1-\inp{u_{n\bm{k}}}{u_{n\bm{k}}}).
\end{equation}

Let's calculate the unconstrained derivatives of $F$ with respect to $\bra{u_{n\bm{k}}}$. 
We first show that 
\begin{align}
  \frac{\delta \mathrm{Tr}[\hat{\rho}_{\bm{k}_2}\hat{o}_{\bm{k}_2}]}{\delta\bra{u_{n\bm{k}_1}}}=&\frac{\delta}{\delta\bra{u_{n\bm{k}_1}}}\sum_{m}\bra{u_{n\bm{k}_2}}\hat{o}_{\bm{k}_2}\ket{u_{n\bm{k}_2}}\nonumber\\
  =&\delta_{\bm{k}_1\bm{k}_2}\hat{o}_{\bm{k}_2}\ket{u_{n\bm{k}_2}}.
\end{align}
The single-particle, Hartree, and Fock energy functionals all take this form thus are easily evaluated
\begin{align}
  \frac{\delta E_{0}}{\delta \bra{u_{n\bm{k}}}}=&\hat{h}^0_{\bm{k}}\ket{u_{n\bm{k}}},\\
  \frac{\delta E_{H}}{\delta\bra{u_{n\bm{k}}}}=&\frac{1}{\mathcal{V}}\sum_{\bm{k}_2\in BZ,\bm{G}}V(\bm{G})\mathrm{Tr}[\hat{\rho}_{\bm{k}_2}\re^{-i\bm{G}\cdot\hat{\bm{r}}}]\re^{i\bm{G}\cdot\hat{\bm{r}}}\ket{u_{n\bm{k}}},\\
  \frac{\delta E_F}{\delta\bra{u_{n\bm{k}}}}=&-\frac{1}{\mathcal{V}}\sum_{\bm{k}_2\in BZ,\bm{G}}V(\bm{k}-\bm{k}_2+\bm{G})\nonumber\\
  &\qquad \qquad \qquad \times\re^{i\bm{G}\cdot\hat{\bm{r}}}\hat{\rho}_{\bm{k}_2}\re^{-i\bm{G\cdot\hat{\bm{r}}}}\ket{u_{n\bm{k}}}.
\end{align}
From the expression above, we could define the Hartree and Fock Hamiltonian as 
\begin{align}
  \hat{h}^H_{\bm{k}}[\hat{\rho}_{\bm{k}}]=&\frac{1}{\mathcal{V}}\sum_{\bm{k}_2\in BZ,\bm{G}}V(\bm{G})\mathrm{Tr}[\hat{\rho}_{\bm{k}_2}\re^{-i\bm{G}\cdot\hat{\bm{r}}}]\re^{i\bm{G}\cdot\hat{\bm{r}}},\label{eq:app-hartree-hamilt}\\
  \hat{h}^F_{\bm{k}}[\hat{\rho}_{\bm{k}}]=&-\frac{1}{\mathcal{V}}\sum_{\bm{k}_2\in BZ,\bm{G}}V(\bm{k}-\bm{k}_2+\bm{G})\re^{i\bm{G}\cdot\hat{\bm{r}}}\hat{\rho}_{\bm{k}_2}\re^{-i\bm{G\cdot\hat{\bm{r}}}}.\label{eq:app-fock-hamilt}
\end{align}
As functionals of gauge invariant single-particle density operator $\hat{\rho}_{\bm{k}}$, the Hartree and Fock Hamiltonian defined in Eq. \eqref{eq:app-hartree-hamilt}\eqref{eq:app-fock-hamilt} are also invariant under $k$-space gauge transform of the occupied bands.

As for the polarization term, we start from the discrete form Eq. \eqref{eq:pol-energy-discrete} and take the thermodynamic limit later.
The unconstrained derivatives of $E_{P}$ is
\begin{align}
  \frac{\delta E_{P}}{\delta\bra{u_{n\bm{k}}}}=&\frac{-e\mathcal{E}}{2i\Delta k_{\parallel}}\frac{\delta}{\delta\bra{u_{n\bm{k}}}}\Big[\sum_{\sigma=\pm}\sigma\sum_{\bm{k}}\log\det S(\bm{k},\bm{k}_{\sigma})\Big]\nonumber\\
  =&\frac{ie\mathcal{E}}{2\Delta k_{\parallel}}\frac{\delta}{\delta\bra{u_{n\bm{k}}}}\Big[\sum_{\sigma=\pm}\sigma\sum_{\bm{k}}\mathrm{Tr} \log S(\bm{k},\bm{k}_{\sigma})\Big]\nonumber\\
  =&\frac{ie\mathcal{E}}{2\Delta k_{\parallel}}\sum_{\sigma=\pm}\sigma\mathrm{Tr}\Big[\frac{\delta S(\bm{k},\bm{k}_{\sigma})}{\delta\bra{u_{n\bm{k}}}}S^{-1}(\bm{k},\bm{k}_{\sigma})\Big]\nonumber\\
  =&\frac{ie\mathcal{E}}{2\Delta k_{\parallel}}\sum_{\sigma=\pm}\sigma\sum_{m=1}^{n_{e}} \ket{u_{m\bm{k}_{\sigma}}}S^{-1}_{mn}(\bm{k},\bm{k}_{\sigma}),
\end{align}
where abbreviation $\bm{k}_{\sigma}=\bm{k}+\sigma\Delta\bm{k}_{\parallel}$ is used for simplicity.
Denote $\ket{D_{n\bm{k}}}=\delta E_{P}/\delta\bra{u_{n\bm{k}}}$.
Easy to see that 
\begin{align}
  \inp{u_{l\bm{k}}}{D_{n\bm{k}}}=&\frac{ie\mathcal{E}}{2\Delta k_{\parallel}}\sum_{\sigma}\sigma\sum_{m=1}^{n_{e}}S_{lm}(\bm{k},\bm{k}_{\sigma})S^{-1}_{mn}(\bm{k},\bm{k}_{\sigma})\nonumber\\
  =&\frac{ie\mathcal{E}}{2\Delta k_{\parallel}}\sum_{\sigma}\sigma\delta_{ln}\nonumber\\
  =&0.
\end{align}
So polarization Hamiltonian could be defined as
\begin{equation}
  \hat{h}^P_{\bm{k}}[\ket{u_{n\bm{k}}};\mathcal{E}]=\sum_{n=1}^{n_{e}}\ket{D_{n\bm{k}}}\bra{u_{n\bm{k}}}+h.c.
\end{equation}
and satisfies
\begin{equation}
  \hat{h}^P_{\bm{k}}\ket{u_{n\bm{k}}}=\sum_{m=1}^{n_{e}}\ket{D_{m\bm{k}}}\delta_{mn}=\ket{D_{n\bm{k}}}=\frac{\delta E_{P}}{\delta\bra{u_{n\bm{k}}}}.
\end{equation}

Before processing, one should verify that this definition of polarization Hamiltonian is a gauge invariant.
By denoting $\Phi^{\dagger}_{\bm{k}}=[\ket{u_{1\bm{k}}},\cdots,\ket{u_{n_{e}\bm{k}}}]$, the polarization Hamiltonian is written in a more neat form 
\begin{equation}
  \hat{h}^P_{\bm{k}}=\frac{ie\mathcal{E}}{2\Delta k_{\parallel}}\sum_{\sigma=\pm}\sigma\Phi_{\bm{k}_{\sigma}}(\Phi^{\dagger}_{\bm{k}}\Phi_{\bm{k}_{\sigma}})^{-1}\Phi^{\dagger}_{\bm{k}}+h.c..
\end{equation}
A $k$-space gauge transformation $(U_{\bm{k}})_{n_{e}\times n_{e}}$ on occupied bands will transform $\Phi_{\bm{k}}$ into $\Phi_{\bm{k}}U_{\bm{k}}$ and the polarization Hamiltonian becomes
\begin{align}
  &(\hat{h}^P_{\bm{k}})'\nonumber\\
  =&\frac{ie\mathcal{E}}{2\Delta k_{\parallel}}\sum_{\sigma=\pm}\sigma\Phi_{\bm{k}_{\sigma}}U_{\bm{k}_{\sigma}}(U^{\dagger}_{\bm{k}}\Phi^{\dagger}_{\bm{k}}\Phi_{\bm{k}_{\sigma}}U_{\bm{k}_{\sigma}})^{-1}U^{\dagger}_{\bm{k}}\Phi^{\dagger}_{\bm{k}}+h.c.\nonumber\\
  =&\frac{ie\mathcal{E}}{2\Delta k_{\parallel}}\sum_{\sigma=\pm}\sigma\Phi_{\bm{k}_{\sigma}}U_{\bm{k}_{\sigma}}U^{-1}_{\bm{k}_{\sigma}}(\Phi^{\dagger}_{\bm{k}}\Phi_{\bm{k}_{\sigma}})^{-1}(U^{\dagger}_{\bm{k}})^{-1}U^{\dagger}_{\bm{k}}\Phi^{\dagger}_{\bm{k}}+h.c.\nonumber\\
  =&\hat{h}^{P}_{\bm{k}},
\end{align}
which is invariant.

It's easier to see this gauge invariance in the thermodynamic limit $L\to \infty$ and $\rd k=\Delta k_{\parallel}\to 0$.
In this limit
\begin{subequations}
  \begin{gather}
    S_{mn}(\bm{k},\bm{k}_{\sigma})=\delta_{mn}+\sigma\inp{u_{m\bm{k}}}{\partial_{k_{\parallel}}u_{n{\bm{k}}}}\rd k,\\
    S^{-1}_{mn}(\bm{k},\bm{k}_{\sigma})=\delta_{mn}-\sigma\inp{u_{m\bm{k}}}{\partial_{k_{\parallel}}u_{n{\bm{k}}}}\rd k.
  \end{gather}
\end{subequations}
So
\begin{align}
  \ket{D_{n\bm{k}}}=&\frac{ie\mathcal{E}}{2\rd k}\sum_{\sigma=\pm}\sigma\sum_{m=1}^{n_{e}}(\ket{u_{m\bm{k}}}+\sigma\ket{\partial_{k_{\parallel}}u_{m\bm{k}}}\rd k)\nonumber\\
  &\quad\times(\delta_{mn}-\sigma\inp{u_{m\bm{k}}}{\partial_{k_{\parallel}}u_{n\bm{k}}}\rd k)\nonumber\\
  =&ie\mathcal{E}\sum_{m=1}^{n_{e}}[\ket{\partial_{k_{\parallel}}u_{m\bm{k}}}\delta_{mn}-\ket{u_{m\bm{k}}}\inp{u_{m\bm{k}}}{\partial_{k_{\parallel}}u_{n\bm{k}}}]\nonumber\\
  =&ie\mathcal{E}(1-\hat{\rho}_{\bm{k}})\ket{\partial_{k_{\parallel}}u_{n\bm{k}}}
\end{align}
and the polarization Hamiltonian in the thermodynamic limit is written as
\begin{align}
  \lim_{\rd k\to 0}\hat{h}^{P}_{\bm{k}}=&ie\mathcal{E}\sum_{n}^{N_{e}}(1-\hat{\rho}_{\bm{k}})\ket{\partial_{{k}_{\parallel}}u_{n\bm{k}}}\bra{u_{n\bm{k}}}+h.c.\nonumber\\
  =&ie\mathcal{E}(1-\hat{\rho}_{\bm{k}})\partial_{k_{\parallel}}\hat{\rho}_{\bm{k}}+h.c.\nonumber\\
  =&ie\bm{\mathcal{E}}\cdot[\nabla_{\bm{k}}\hat{\rho}_{\bm{k}},\hat{\rho}_{\bm{k}}]. \label{eq:hp-thermodynamic}
\end{align}
The thermodynamic limit expression Eq. \eqref{eq:hp-thermodynamic} is only a functional of the gauge invariant $\hat{\rho}_{\bm{k}}$ and thus is also a gauge invariant.

Finally, minimization of $F[\ket{u_{n\bm{k}}};\mathcal{E}]$ gives us the self-consistent equation
\begin{equation}
  \frac{\delta F}{\delta\bra{u_{n\bm{k}}}}=0\implies \hat{h}^{MF}_{\bm{k}}[\ket{u_{n\bm{k}}};\mathcal{E}]\ket{u_{n\bm{k}}}=\xi_{n\bm{k}}\ket{u_{n\bm{k}}}
\end{equation}
where the mean-field Hamiltonian is
\begin{equation}
  \hat{h}^{MF}_{\bm{k}}=\hat{h}^0_{\bm{k}}+\hat{h}^{H}_{\bm{k}}[\hat{\rho}_{\bm{k}}]+\hat{h}^F_{\bm{k}}[\hat{\rho}_{\bm{k}}]+\hat{h}^P_{\bm{k}}[\ket{u_{n\bm{k}}};\mathcal{E}].
\end{equation}

\section{The Hessian Matrix}\label{app:hessian}
Assume $\mathcal{E}<\mathcal{E}_c$, and the self consistent equation has solutions
\begin{equation}
  h^{MF}_{\bm{k}}[\ket{v\bm{k};\mathcal{E}}]\ket{i\bm{k};\mathcal{E}} =\xi_{i\bm{k};\mathcal{E}}\ket{i\bm{k};\mathcal{E}},\quad i=c,v .
\end{equation}
The valence band $\ket{v\bm{k};\mathcal{E}}$ is chosen as the one with lower band energy, i.e. $\xi_{v\bm{k};\mathcal{E}}<\xi_{c\bm{k};\mathcal{E}}$,
The $\mathcal{E}$ label in wavefunctions and band energies means they are converged solutions.

At the converged point (local minimum of the total energy functional), the trial HF state could be reparameterized as 
\begin{equation}
  \ket{v'\bm{k};\mathcal{E}}=\frac{\ket{v\bm{k};\mathcal{E}}+f_{\bm{k}}\ket{c\bm{k};\mathcal{E}}}{\sqrt{1+|f_{\bm{k}}|^2}},\label{eq:reparameterization_app}
\end{equation}
where $f_{\bm{k}}$ is arbitrary complex-valued function defined on Brillouin zone.
This parametrization is unconstrained and complete, and the total energy then becomes functional of $f_{\bm{k}}$ as 
\begin{equation}
  \varepsilon_{tot}[f_{\bm{k}}^*,f_{\bm{k}};\mathcal{E}]\equiv \varepsilon_{tot}[\ket{v'\bm{k};\mathcal{E}};\mathcal{E}].
\end{equation}
By writing $f_{\bm{k}}=f_{\bm{k},r}+if_{\bm{k},i}$, where $f_{\bm{k},r}$ and $f_{\bm{k},i}$ are real variables, the Hessian matrix is defined as 
\begin{equation}
  \mathrm{H}_{\bm{k}\bm{k}'}=\begin{pmatrix}
    \frac{\delta^2 \varepsilon_{tot}}{\delta f_{\bm{k},r}\delta f_{\bm{k}',r}} & \frac{\delta^2 \varepsilon_{tot}}{\delta f_{\bm{k},r}\delta f_{\bm{k}',i}}\\
    \frac{\delta^2 \varepsilon_{tot}}{\delta f_{\bm{k},i}\delta f_{\bm{k}',r}} & \frac{\delta^2 \varepsilon_{tot}}{\delta f_{\bm{k},i}\delta f_{\bm{k}',i}}
  \end{pmatrix}.
\end{equation}

For simplicity, the $\mathcal{E}$ label will be omitted in the following derivations.

We first calculate the derivatives of $\ket{v'\bm{k} }$ with respect to $f_{\bm{k},r/i}$ for further usage.
\begin{gather}
  \frac{\delta\ket{v'\bm{k}}}{\delta f_{\bm{k},r}}=\frac{-f_{\bm{k},r}\ket{v\bm{k} }+(1-if_{\bm{k},i}f_{\bm{k}})\ket{c\bm{k} }}{(1+|f_{\bm{k}}|^2)^{3/2}},\\
  \frac{\delta\ket{v'\bm{k} }}{\delta f_{\bm{k},i}}=\frac{-f_{\bm{k},i}\ket{v\bm{k} }+i(1+f_{\bm{k},r}f_{\bm{k}})\ket{c\bm{k} }}{(1+|f_{\bm{k}}|^2)^{3/2}}.
\end{gather}
At $f_{\bm{k}}=0$, they are simplified as 
\begin{gather}
  \frac{\delta\ket{v'\bm{k} }}{\delta f_{\bm{k},r}}\Big|_{f_{\bm{k}}=0}=\ket{c\bm{k} },\;
\frac{\delta\ket{v'\bm{k} }}{\delta f_{\bm{k},i}}\Big|_{f_{\bm{k}}=0}=i\ket{c\bm{k} }.
\end{gather}
The second order derivatives of $\ket{v'\bm{k} }$ at $f_{\bm{k}}=0$ are 
\begin{gather}
  \frac{\delta^2 \ket{v'\bm{k} }}{\delta f_{\bm{k},r}\delta f_{\bm{k}',r}}\Big|_{f_{\bm{k}}=0}=-\delta_{\bm{k}\bm{k}'}\ket{v\bm{k} },\\
  =\frac{\delta^2 \ket{v'\bm{k} }}{\delta f_{\bm{k},i}\delta f_{\bm{k}',i}}\Big|_{f_{\bm{k}}=0}=-\delta_{\bm{k}\bm{k}'}\ket{v\bm{k} },\\
  \frac{\delta^2 \ket{v'\bm{k} }}{\delta f_{\bm{k},r}\delta f_{\bm{k}',i}}\Big|_{f_{\bm{k}}=0}=0.
\end{gather}

The first order derivative of $\varepsilon_{tot}$ defined by Eq. \eqref{eq:tot-energy} is
\begin{align}
  \frac{\delta \varepsilon_{tot}}{\delta f_{\bm{k},r/i}}=&\frac{\delta \bra{v'\bm{k} }}{\delta f_{\bm{k},r/i}}\frac{\delta \varepsilon_{tot}}{\delta \bra{v'\bm{k} }}+c.c.\nonumber\\
  =&\frac{1}{\mathcal{V}}\frac{\delta \bra{v'\bm{k} }}{\delta f_{\bm{k},r/i}}h_{\bm{k}}^{MF}[\ket{v'\bm{k} }]\ket{v'\bm{k} }+c.c..\label{eq:first_order}
\end{align}
We use the definition of mean-field Hamiltonian $h^{MF}_{\bm{k}}[\ket{v\bm{k}}]\ket{v\bm{k}}\equiv \mathcal{V}\delta \varepsilon_{tot}/\delta \bra{v\bm{k}}$ for the last equality in Eq. \eqref{eq:first_order}.
At $f_{\bm{k}}=0$, the first-order derivative is just
\begin{equation}
  \frac{\delta \varepsilon_{tot}}{\delta f_{\bm{k},r/i}}\Big|_{f_{\bm{k}}=0}\propto\bra{c\bm{k} }h^{MF}_{\bm{k}}[\ket{v\bm{k} }]\ket{v\bm{k} }+c.c.=0,\nonumber
\end{equation}
which is consistent with the fact that $\ket{v\bm{k} }$ is a local minimum.

Then let's evaluate second-order derivatives of $\varepsilon_{tot}$
\begin{align}
  &\mathcal{V}\frac{\delta^2\varepsilon_{tot}}{\delta f_{\bm{k},r}\delta {f_{\bm{k}',r}}}\Big|_{f_{\bm{k}}=0}\nonumber\\
  =&\frac{\delta \bra{v'\bm{k} }}{\delta f_{\bm{k},r}}\Big|_{f_{\bm{k}}=0}\frac{\delta h_{\bm{k}}^{MF}[\ket{v'\bm{k} }]}{\delta f_{\bm{k}',r}}\Big|_{f_{\bm{k}}=0}\ket{v'\bm{k} }\nonumber\\
  &+\frac{\delta \bra{v'\bm{k} }}{\delta f_{\bm{k},r}}\Big|_{f_{\bm{k}}=0}h^{MF}_{\bm{k}}[\ket{v'\bm{k} }]\frac{\delta \ket{v'\bm{k} }}{\delta f_{\bm{k}',r}}\Big|_{f_{\bm{k}}=0}\nonumber\\
  &+\frac{\delta^2\bra{v'\bm{k} }}{\delta f_{\bm{k},r}\delta f_{\bm{k}',r}}\Big|_{f_{\bm{k}}=0}h_{\bm{k}}^{MF}[\ket{v'\bm{k} }]\ket{v'\bm{k} }+c.c.\nonumber\\
  =&\delta_{\bm{k}\bm{k}'}(\xi_{c\bm{k} }-\xi_{v\bm{k} })+\bra{c\bm{k} }\frac{\delta h_{\bm{k}}^{MF}[\ket{v'\bm{k} }]}{\delta f_{\bm{k}',r}}\Big|_{f_{\bm{k}}=0}\ket{v\bm{k} }+c.c..\nonumber
\end{align}
Similarly,
\begin{align}
  &\mathcal{V}\frac{\delta^2\varepsilon_{tot}}{\delta f_{\bm{k},i}\delta {f_{\bm{k}',i}}}\Big|_{f_{\bm{k}}=0}\nonumber\\
  =&\delta_{\bm{k}\bm{k}'}(\xi_{c\bm{k} }-\xi_{v\bm{k} })-i\bra{c\bm{k} }\frac{\delta h_{\bm{k}}^{MF}[\ket{v'\bm{k} }]}{\delta f_{\bm{k}',i}}\Big|_{f_{\bm{k}}=0}\ket{v\bm{k} }+c.c.\nonumber
\end{align}
and
\begin{align}
  \mathcal{V}\frac{\delta^2\varepsilon_{tot}}{\delta f_{\bm{k},r}\delta {f_{\bm{k}',i}}}\Big|_{f_{\bm{k}}=0}=\bra{c\bm{k} }\frac{\delta h_{\bm{k}}^{MF}[\ket{v'\bm{k} }]}{\delta f_{\bm{k}',i}}\Big|_{f_{\bm{k}}=0}\ket{v\bm{k} }+c.c..\nonumber
\end{align}
So the final task is to evaluate the derivatives of $h^{MF}_{\bm{k}}$ with respect to $f_{\bm{k},r/i}$.

The Hartree one is
\begin{align}
  &\bra{c\bm{k}}\frac{\delta h^H}{\delta f_{\bm{k}',r}}\Big|_{f_{\bm{k}}=0}\ket{v\bm{k}}\nonumber\\
  =&\frac{4\pi e^2 d\inp{c\bm{k}}{e}\inp{e}{v\bm{k}}}{\epsilon}\frac{\delta n_{ex}}{\delta f_{\bm{k}',r}}\Big|_{f_{\bm{k}}=0}\nonumber\\
  =&\frac{4\pi e^2 d\inp{c\bm{k}}{e}\inp{e}{v\bm{k}}}{\epsilon}\left[\frac{\delta \bra{v'\bm{k} }}{\delta f_{\bm{k}',r}}\frac{\delta n_{ex}}{\delta\bra{v'\bm{k} }}\Big|_{f_{\bm{k}}=0}+c.c.\right]\nonumber\\
  =&\frac{4\pi e^2 d\inp{c\bm{k}}{e}\inp{e}{v\bm{k}}}{\epsilon}\left[\frac{1}{\mathcal{V}}\inp{c\bm{k}' }{e}\inp{e}{v\bm{k}' }+c.c.\right]\nonumber\\
  =&\frac{2}{\mathcal{V}}\frac{4\pi e^2 d\inp{c\bm{k}}{e}\inp{e}{v\bm{k}}}{\epsilon}\mathrm{Re}[\inp{c\bm{k}' }{e}\inp{e}{v\bm{k}' }]\nonumber
\end{align}
and 
\begin{align}
  &\bra{c\bm{k}}\frac{\delta h^H}{\delta f_{\bm{k}',i}}\Big|_{f_{\bm{k}}=0}\ket{v\bm{k}}\nonumber\\
  =&\frac{4\pi e^2 d\inp{c\bm{k}}{e}\inp{e}{v\bm{k}}}{\epsilon}\left[-i\frac{1}{\mathcal{V}}\inp{c\bm{k}' }{e}\inp{e}{v\bm{k}' }+c.c.\right]\nonumber\\
  =&\frac{2}{\mathcal{V}}\frac{4\pi e^2 d\inp{c\bm{k}}{e}\inp{e}{v\bm{k}}}{\epsilon}\mathrm{Im}[\inp{c\bm{k}' }{e}\inp{e}{v\bm{k}' }].\nonumber
\end{align}
The Fock one is
\begin{align}
  &\bra{c\bm{k}}\frac{\delta h^F_{\bm{k}}}{\delta f_{\bm{k}',r}}\Big|_{f_{\bm{k}}=0}\ket{v\bm{k}}\nonumber\\
  =&-\frac{1}{\mathcal{V}}\sum_{ss'}V_{s's}(\bm{k}-\bm{k}')\inp{c\bm{k}}{s}\inp{s'}{v\bm{k}}\frac{\delta \rho_{ss'\bm{k}'}}{\delta f_{\bm{k}',r}}\Big|_{f_{\bm{k}}=0}\nonumber\\
  =&-\frac{1}{\mathcal{V}}\sum_{ss'}V_{s's}(\bm{k}-\bm{k}')\inp{c\bm{k}}{s}\inp{s'}{v\bm{k}}\nonumber\\
  &\qquad \qquad \times(\inp{v\bm{k}'}{s'}\inp{s}{c\bm{k}'}+\inp{c\bm{k}'}{s'}\inp{s}{v\bm{k}'})\nonumber
\end{align}
and 
\begin{align}
  &\bra{c\bm{k}}\frac{\delta h^F_{ss'\bm{k}}}{\delta f_{\bm{k}',i}}\Big|_{f_{\bm{k}}=0}\ket{v\bm{k}}\nonumber\\
  =&-\frac{i}{\mathcal{V}}\sum_{ss'}V_{s's}(\bm{k}-\bm{k}')\inp{c\bm{k}}{s}\inp{s'}{v\bm{k}}\nonumber\\
  &\qquad \qquad \times(\inp{v\bm{k}'}{s'}\inp{s}{c\bm{k}'}-\inp{c\bm{k}'}{s'}\inp{s}{v\bm{k}'}).\nonumber
\end{align}
As for the polarization term, use the fact that
\begin{equation}
  \bra{c\bm{k}}h^{P}_{\bm{k}}\frac{\delta \ket{v'\bm{k}}}{\delta f_{\bm{k}'}}\Big|_{f_{\bm{k}}=0}\propto\delta_{\bm{k}\bm{k}'}\bra{c\bm{k}}h^P_{\bm{k}}\ket{c\bm{k}}=0,\nonumber
\end{equation}
we have 
\begin{align}
  &\bra{c\bm{k}}\frac{\delta h^P_{\bm{k}}}{\delta f_{\bm{k}',r}}\Big|_{f_{\bm{k}}=0}\ket{v\bm{k}}=\bra{c\bm{k}}\frac{\delta (h^P_{\bm{k}}\ket{v'\bm{k}})}{\delta f_{\bm{k}',r}}\Big|_{f_{\bm{k}}=0}\nonumber\\
  =&\bra{c\bm{k}}\frac{\delta}{\delta f_{\bm{k}',r}}\left[\frac{ie\mathcal{E}}{2\Delta k_{\parallel}}\sum_{\sigma=\pm 1}\frac{\sigma\ket{v'\bm{k}_{\sigma}}}{\inp{v'\bm{k}}{v'\bm{k}_{\sigma}}}\right]\Big|_{f_{\bm{k}}=0}\nonumber\\
  =&\frac{ie\mathcal{E}}{2\Delta k_{\parallel}}\sum_{\sigma=\pm}\sigma\left[\delta_{\bm{k}'\bm{k}_{\sigma}}\frac{\inp{c\bm{k}}{c\bm{k}_{\sigma}}}{\inp{v\bm{k}}{v\bm{k}_{\sigma}}}-\delta_{\bm{k}'\bm{k}}\frac{(\inp{c\bm{k}}{v\bm{k}_{\sigma}})^2}{(\inp{v\bm{k}}{v\bm{k}_{\sigma}})^2}\right.\nonumber\\ 
  &\left. -\delta_{\bm{k}'\bm{k}_{\sigma}}\frac{\inp{c\bm{k}}{v\bm{k}_{\sigma}}\inp{v\bm{k}}{c\bm{k}_{\sigma}}}{(\inp{v\bm{k}}{v\bm{k}_{\sigma}})^2}\right]\nonumber
\end{align}
and 
\begin{align}
  &\bra{c\bm{k}}\frac{\delta h^P_{\bm{k}}}{\delta f_{\bm{k}',i}}\Big|_{f_{\bm{k}}=0}\ket{v\bm{k}}\nonumber\\
  =&\frac{ie\mathcal{E}}{2\Delta k_{\parallel}}\sum_{\sigma=\pm}\sigma\left[i\delta_{\bm{k}'\bm{k}_{\sigma}}\frac{\inp{c\bm{k}}{c\bm{k}_{\sigma}}}{\inp{v\bm{k}}{v\bm{k}_{\sigma}}}+i\delta_{\bm{k}'\bm{k}}\frac{(\inp{c\bm{k}}{v\bm{k}_{\sigma}})^2}{(\inp{v\bm{k}}{v\bm{k}_{\sigma}})^2}\right.\nonumber\\ 
  &\left. -i\delta_{\bm{k}'\bm{k}_{\sigma}}\frac{\inp{c\bm{k}}{v\bm{k}_{\sigma}}\inp{v\bm{k}}{c\bm{k}_{\sigma}}}{(\inp{v\bm{k}}{v\bm{k}_{\sigma}})^2}\right].\nonumber
\end{align}

\section{The Goldstone Mode}\label{app:Goldstone_mode}
The many-body Hamiltonian Eq. \eqref{eq:manybody_hamiltonian} is invariant under gauge transformations of the electron creation operators: $c^{\dagger}_{e\bm{k}}\to \re^{i\phi_{e}}c^{\dagger}_{e\bm{k}}$, $c^{\dagger}_{h\bm{k}}\to \re^{i\phi_{h}}c^{\dagger}_{h\bm{k}}$.
This $U(1)\times U(1)$ symmetry corresponds to the charge conservation in each layer.

After this gauge transformation, the valance band electron creation operator becomes 
\begin{equation}
  (c^{\dagger}_{v\bm{k}})'= \alpha_{\bm{k}}\re^{i\phi_{e}}c^{\dagger}_{e\bm{k}}+\beta_{\bm{k}}\re^{i\phi_{h}}c^{\dagger}_{h\bm{k}},
\end{equation}
which gives a new trial wavefunction
\begin{equation}
  \ket{v'\bm{k}}=\begin{bmatrix}
    \re^{i\phi_e}\alpha_{\bm{k}}\\
    \re^{i\phi_h}\beta_{\bm{k}}
  \end{bmatrix}=\re^{i\phi}\begin{bmatrix}
    \re^{i\phi_{ex}}\alpha_{\bm{k}}\\
    \re^{-i\phi_{ex}}\beta_{\bm{k}}
  \end{bmatrix},\label{eq:transform_wave}
\end{equation}
where $\phi=(\phi_e+\phi_h)/2$, $\phi_{ex}=(\phi_e-\phi_h)/2$ are related to the conservation of total charge and exciton number respectively.
The relative density matrix $\tilde{\rho}=\rho-\rho^0$ transforms into 
\begin{equation}
\tilde{\rho}'_{\bm{k}}=\begin{bmatrix}
    |\alpha_{\bm{k}}|^2 & \re^{i(\phi_e-\phi_h)}\alpha_{\bm{k}}\beta^*_{\bm{k}} \\
    \re^{i(\phi_h-\phi_e)}\alpha^*_{\bm{k}}\beta_{\bm{k}} & |\beta_{\bm{k}}|-1
  \end{bmatrix},\label{eq:transform_dst_mat}
\end{equation}
or equivalently, $\tilde{\rho}'_{ss'\bm{k}}= \re^{i(\phi_{s}-\phi_{s'})}\tilde{\rho}_{ss'\bm{k}}$.
Besides, the overlap matrix $S(\bm{k},\bm{k})=\inp{v\bm{k}}{v\bm{k}'}$ becomes
\begin{equation}
  S'(\bm{k},\bm{k}')=\inp{v'\bm{k}}{v'\bm{k}'}=\alpha^*_{\bm{k}}\alpha_{\bm{k}'}+\beta^*_{\bm{k}}\beta_{\bm{k}'}=S(\bm{k},\bm{k}').\label{eq:transform_overlap_mat}
\end{equation}
Substitute Eq. \eqref{eq:transform_dst_mat}\eqref{eq:transform_overlap_mat} into the total energy expression Eq. \eqref{eq:tot-energy} we find that $\varepsilon_{tot}[\ket{v'\bm{k}};\mathcal{E}]=\varepsilon_{tot}[\ket{v\bm{k}};\mathcal{E}]$, i.e. the total energy is invariant under the transformation $\ket{v\bm{k}}\to \ket{v'\bm{k}}$.

The $U(1)$ symmetry related to exciton conservation (phase $\phi_{ex}=\phi_e-\phi_h$ of electron-hole pairing condensate $\rho_{eh\bm{k}}$) gives a zero energy Goldstone mode to the valance band fluctuation.
To see this, let's rewrite $\ket{v'\bm{k}}$ into a linear combination of $\ket{v\bm{k}}=[\alpha_{\bm{k}},\beta_{\bm{k}}]^T$ and $\ket{c\bm{k}}=\re^{-i\varphi_{\bm{k}}}[-\beta^*_{\bm{k}},\alpha^*_{\bm{k}}]^T$ ($\varphi_{\bm{k}}$ is an arbitrary function of $\bm{k}$) as
\begin{align}
  \re^{-i\phi}\ket{v'\bm{k}} =&\inp{v\bm{k}}{v'\bm{k}}\ket{v\bm{k}}+\inp{c\bm{k}}{v'\bm{k}}\ket{c\bm{k}}\nonumber\\
  =&(\re^{i\phi_{ex}}|\alpha_{\bm{k}}|^2+\re^{-i\phi_{ex}}|\beta_{\bm{k}}|^2)\ket{v\bm{k}}\nonumber\\
  &\quad -2i\alpha_{\bm{k}}\beta_{\bm{k}}\sin\phi_{ex}\re^{i\varphi_{\bm{k}}}\ket{c\bm{k}}\nonumber\\
  \sim&\ket{v\bm{k}}-2i\phi_{ex}\alpha_{\bm{k}}\beta_{\bm{k}}\re^{i\varphi_{\bm{k}}}\ket{c\bm{k}},\label{eq:Goldstone_mode}
\end{align}
Compare Eq. \eqref{eq:reparameterization_app}\eqref{eq:Goldstone_mode} we find the corresponding Goldstone mode in parameter space is just expressed as
\begin{equation}
  f^{Gs}_{\bm{k}}\propto i\alpha_{\bm{k}}\beta_{\bm{k}}\re^{i\varphi_{\bm{k}}}=i\inp{e}{v\bm{k};\mathcal{E}}\inp{c\bm{k};\mathcal{E}}{e}.\label{eq:Goldstone_mode_parameterization}
\end{equation}

\begin{figure}
  \centering
  \includegraphics[width=0.5\linewidth]{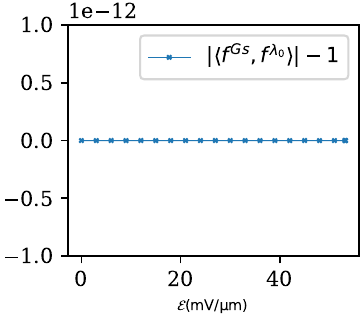}
  \caption{Overlap between the Goldstone mode Eq. \eqref{eq:Goldstone_mode_parameterization} and the zero mode of the Hessian matrix Eq. \eqref{eq:Hessian}.}
  \label{fig:overlap}
\end{figure}
The overlap between the Goldstone mode $f^{Gs}_{\bm{k}}$ and the zero mode $f^{\lambda_0}_{\bm{k}}$ of the Hessian matrix Eq. \eqref{eq:Hessian} is calculated as
\begin{equation}
  I=\big|\langle f^{Gs}_{\bm{k}},f^{\lambda_0}_{\bm{k}}\rangle\big|=\Big|\sum_{\bm{k}}(f^{Gs}_{\bm{k}})^*f^{\lambda_0}_{\bm{k}}\Big|
\end{equation}
and plotted in Fig. \ref{fig:overlap}.
The results show that $I$ is equal to 1 in numerical precision, which means the zero mode of the Hessian matrix is indeed the Goldstone mode $f^{Gs}_{\bm{k}}$ discussed in this section.

\section{The Inter-band Zener Tunneling}\label{app:zener_tunneling}

Consider the inter-band tunneling problem of the 2D continuous model 
\begin{equation}
  \hat{h}=\begin{bmatrix}
    -\frac{\partial_{x}^2}{2m}-\frac{\partial_{y}^2}{2m}-\frac{\mu^0_{ex}}{2} & \frac{\Delta}{2} \\
    \frac{\Delta}{2} & \frac{\partial_{x}^2}{2m}+\frac{\partial_{y}^2}{2m}+\frac{\mu^0_{ex}}{2}
  \end{bmatrix}+V(x)
\end{equation}
where the barrier potential $V(x)$ is defined as 
\begin{equation}
  V(x)=\left\{\begin{aligned}
      e\mathcal{E}L/2,& \; x\le -L/2 \\
    -e\mathcal{E}x, & \; -L/2\le x\le L/2\\
    -e\mathcal{E}L/2, & \; x\ge L/2
  \end{aligned}\right.
\end{equation}
For a given tunneling energy $E$, the Schr\"odinger is 
\begin{equation}
  \hat{h}\ket{\Psi;E}=E\ket{\Psi;E}.\label{eq:sch_eq}
\end{equation}
Since the electrical field is applied only along $x$-direction, translation symmetry in $y$ direction still holds and $k_y$ is a good quantum number.
Following Zener\cite{zenerTheoryElectricalBreakdown1934}, we could write the approximated WKB wavefunction as
\begin{equation}
  \ket{\Psi_{k_y};E}\propto \exp\left[ik_y y+i\int_{-\infty}^{x}k(x')\rd x'\right]\ket{\tilde{u}_{k(x)k_y}}.\label{eq:wkb_wave}
\end{equation}
If $k(x)$ is slow varying so that $\partial_{x}k(x)$ could be neglected, substitute Eq. \eqref{eq:wkb_wave} into the Schr\"odinger equation we find that 
\begin{equation}
  h_{k(x)k_y}\ket{\tilde{u}_{k(x)k_y}}=(E-V(x))\ket{\tilde{u}_{k(x)k_y}},\label{eq:secular_eq}
\end{equation}
where
\begin{equation}
  h_{k(x)k_y}=\begin{bmatrix}
    \frac{k^2(x)}{2m}-\frac{\mu_{ex}(k_y)}{2} & \frac{\Delta}{2} \\
    \frac{\Delta}{2} & -\frac{k^2(x)}{2m}+\frac{\mu_{ex}(k_y)}{2}.
  \end{bmatrix}
\end{equation}
and $\mu_{ex}(k_y)=\mu^0_{ex}-k_y^2/m$.
Solving the secular equation \eqref{eq:secular_eq} gives the relation between the complex wavevector $k(x)$ and position $x$
\begin{equation}
  [k^2(x)-m\mu_{ex}]^2+(m\Delta)^2=[2m(E-V(x))]^2\label{eq:vector_position}
\end{equation}
Things are different for $\mu_{ex}>0$ and $\mu_{ex}<0$ and should be discussed separately.
The condition $\mu_{ex}(k_y)=0$ gives a critical $k_y$ as 
\begin{equation}
  \mu_{ex}(k_y)=\mu_{ex}^0-k_y^2/m=0\implies k_{y,c}=\sqrt{m\mu_{ex}^0}.
\end{equation}

\begin{figure}[!hbpt]
  \centering
  \def\svgwidth{\linewidth}
  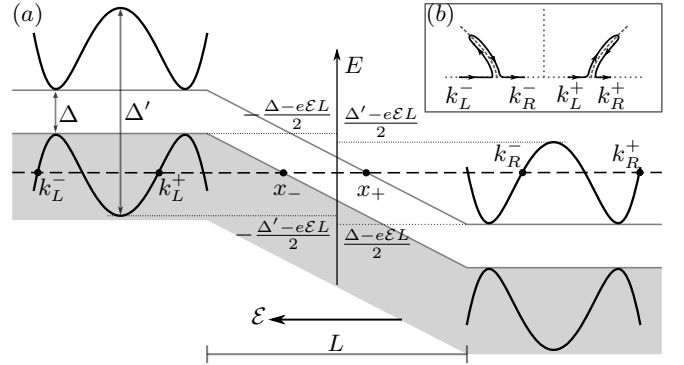
  \caption{(a) Tunneling scenario for $\mu_{ex}(k_y)>0$. 
  The tunneling channels $k_{L}^{\pm}\to k_{R}^{\pm}$ only exist when in-plane bias voltage overcomes the band gap, i.e. $e\mathcal{E}L>\Delta$. 
  Under WKB approximation, the valance band $k_{L}^{\sigma}$ states in the region $x\le -L/2$ will continuously turn into the conduction band $k_{R}^{\sigma}$ states in the region $x\ge L/2$ as propagating to the right. 
  $x_{\pm}=(\pm\Delta/2-E)/e\mathcal{E}$ marks the classical turning points. 
  (b) The paths of the complex wavevectors $k^{\sigma}(x)$ in the complex plane are indicated by the black arrow lines.}
  \label{fig:tunnel_band_inversion}
\end{figure}

The tunneling scenario  for $\mu_{ex}(k_y)>0$ (or equivalently $k_y^2\le k_{y,c}^{2}$) is illustrated in Fig. \ref{fig:tunnel_band_inversion}(a).
As a tunneling state propagating to the right, $\ket{\Psi_{k_y};E}$ should behave like a valance band electron in the region $x\ll -L/2$ ($k_{L}^{\pm}$ states in Fig. \ref{fig:tunnel_band_inversion}(a)) and like a conduction band electron in the region $x\gg L/2$ ($k_{R}^{\pm}$ states in Fig. \ref{fig:tunnel_band_inversion}(a)).
This places a restriction on the tunneling energy $(\Delta-e\mathcal{E}L)/2\le E \le -(\Delta-e\mathcal{E}L)/2$, which further demands that $e\mathcal{E}L\ge \Delta$.
In other words, the inter-band Zener tunneling only occurs when the in-plane bias voltage exceeds the band gap.

As the electron propagates to the right in the region $|x|\le L/2$, the complex wavevector $k(x)$ will travel from $k_{L}^{\sigma}$ to $k_{R}^{\sigma}$ in the complex plane along the line\cite{kaneInterbandTunneling1969}
\begin{equation}
  \mathrm{Im}[k^2(x)-m\mu_{ex}]^2=0.\label{eq:vector_position_im}
\end{equation}
Eq. \eqref{eq:vector_position_im} is just the imaginary part of Eq. \eqref{eq:vector_position} and is solved as 
\begin{equation}
  \mathrm{Im}k\times\mathrm{Re}k\times[(\mathrm{Re}k)^2-(\mathrm{Im}k)^2-m\mu_{ex}]=0.\label{eq:vector_path}
\end{equation}
The solutions of Eq. \eqref{eq:vector_path} in the complex plane are represented by dashed gray lines in Fig. \ref{fig:tunnel_band_inversion}(b).
The paths of $k^{\sigma}(x)$ in the complex plane are also illustrated by solid black arrow lines in Fig. \ref{fig:tunnel_band_inversion}(b).
This analysis means that $k^{+}(x)$ and $k^{-}(x)$ are two independent tunneling channels.

Need to notice that, the tunneling channel $k^{+}(x)$ only exists for tunneling energy $E\ge-(\Delta'-e\mathcal{E}L)/2$ where $\Delta'=\sqrt{\mu_{ex}^2+\Delta^2}$.
This is because there is no $k_{L}^{+}$ state in the region $x\ll -L/2$ when $E<-(\Delta'-e\mathcal{E}L)/2$ as is shown in Fig. \ref{fig:tunnel_band_inversion}(a).
So the allowed tunneling energy range for $k^{+}(x)$ channel is $E^{+}_{max}=(e\mathcal{E}L-\Delta)/2$ and $E^{+}_{min}=\max(-(\Delta'-e\mathcal{E}L)/2,(\Delta-e\mathcal{E}L)/2)$.
Similarly, the tunneling channel $k^{-}(x)$ only exists when tunneling energy is in the range $E^{-}_{max}=\min((\Delta'-e\mathcal{E}L)/2,(e\mathcal{E}L-\Delta)/2)$ and $E^{-}_{min}=(\Delta-e\mathcal{E}L)/2$.

Once these energy conditions are satisfied, one can calculate the tunneling probability under WKB approximation directly by
\begin{equation}
  P^{WKB}_{k_{L}^{\sigma}k_{R}^{\sigma},k_y}(E)=\frac{|\Psi_{k_y}(x=L/2;E)|^2}{|\Psi_{k_y}(x=-L/2;E)|^2}=\re^{-2\zeta_{k_y}^{\sigma}(E)},\nonumber
\end{equation}
where $\zeta^{\sigma}_{k_y}(E)$ is the Zener parameter defined by
\begin{equation}
  \zeta^{\sigma}_{k_y}(E)\equiv\int_{x_{-}}^{x_{+}}\rd x\;|\mathrm{Im}k^{\sigma}(x)|
\end{equation}
The lower and upper limits $x_{\pm}=(\pm\Delta/2- E)/e\mathcal{E}$ of the integration are the classical turning points.
Only in the range $x_{-}\le x\le x_{+}$, $k^{\sigma}(x)$ has an imaginary part
\begin{equation}
  |\mathrm{Im}k^{\sigma}(x)|=\sqrt{\frac{m}{2}}\sqrt{\sqrt{\mu^2_{ex}+\Delta^2-4(E+e\mathcal{E}x)^2}-\mu_{ex}}.\nonumber
\end{equation}
So the Zener parameter is calculated as 
\begin{align}
  \zeta_{k_y}^{\sigma}(E)
  =&\frac{\sqrt{m}}{2\sqrt{2}e\mathcal{E}}\int_{-\Delta}^{\Delta}\rd E\;\sqrt{\sqrt{\mu^2_{ex}+\Delta^2-E^2}-\mu_{ex}}\nonumber\\
  =&\frac{\sqrt{m}\Delta^{3/2}}{\sqrt{2}e\mathcal{E}}\int_0^{1}\rd \varepsilon\;\sqrt{\sqrt{\tilde{\mu}^2_{ex}+1-\varepsilon^2}-\tilde{\mu}_{ex}},\label{eq:zener_band_invert}
\end{align}
where $\tilde{\mu}_{ex}=\mu_{ex}(k_y)/\Delta=(\mu^0_{ex}-k_y^2/m)/\Delta>0$.
One can see that $\zeta_{k_y}^{\sigma}(E)=\zeta(k_y)$ is only a function of $k_y$.
So the transition probability is also only a function of $k_y$, i.e. $P^{WKB}_{k_L^{\sigma}k_R^{\sigma},k_y}(E)=P(k_y)=\re^{-2\zeta(k_y)}$.

The current contributed by state $\ket{\Psi_{k_y};k_{L}^{\sigma}\to k_{R}^{\sigma}}$ is calculated by multiplying the tunneling probability with the velocity $v_{c,k_{R}^{\sigma}k_y}=\partial_{k_{R}^{\sigma}}\varepsilon_{c,k_{R}^{\sigma}k_y}$ of the final state.
Sum all possible final states $k_{R}^{\sigma}$ together and we get
\begin{align}
  j(k_y)=&-e\sum_{\sigma}\int\frac{\rd k_{R}^{\sigma}}{2\pi}\;P^{WKB}_{k_L^{\sigma}k_R^{\sigma},k_y}(E)\partial_{k_{R}^{\sigma}}\varepsilon_{c,k_{R}^{\sigma}k_y}\nonumber\\
  =&-\frac{eP(k_y)}{2\pi}\sum_{\sigma}\int_{E^{\sigma}_{min}}^{E^{\sigma}_{max}}\rd E\nonumber\\
  =&-\frac{eP(k_y)}{2\pi}\delta E(k_y),
  \label{eq:current_band_invert}
\end{align}
where $\delta E(k_y)=\min\left(2(e\mathcal{E}L-\Delta),\Delta'-\Delta\right)$.

\begin{figure}[!hbpt]
  \centering
  \def\svgwidth{\linewidth}
  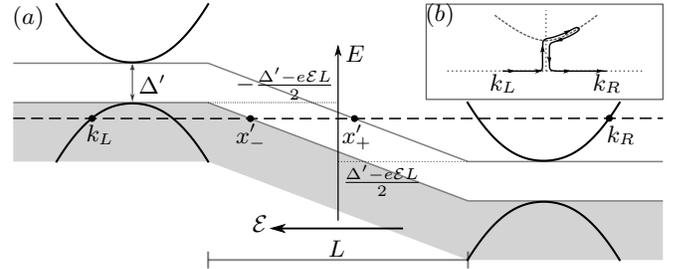
  \caption{(a) Tunneling scenario for $\mu_{ex}(k_y)<0$. In this case, there is no band inversion and the band gap becomes $\Delta'=\sqrt{\Delta^2+\mu^2_{ex}}$. There exists one and only one tunneling channel $k_{L}\to k_{R}$ when tunneling energy satisfies $|E|\le (e\mathcal{E}L-\Delta')/2$. 
  (b) The path of the complex wavevector $k(x)$ in the complex plane is indicated by the black arrow line.}
  \label{fig:tunnel_no_band_inversion}
\end{figure}

On the other hand, the tunneling scenario for the case $\mu_{ex}<0$ (or equivalently $k_y^2>k_{y,c}^2$) is shown in Fig. \ref{fig:tunnel_no_band_inversion}(a).
Different from the case $\mu_{ex}>0$, there exists one and only one tunneling channel $\ket{\Psi_{k_y};k_{L}\to k_R}$ for tunneling energy in the range $E_{min}=(\Delta'-e\mathcal{E}L)/2\le 0$ and $E_{max}=(e\mathcal{E}L-\Delta')/2\ge0$.
And the path of the wavevector $k(x)$ in the complex plane is indicated by the black solid arrow line in Fig. \ref{fig:tunnel_no_band_inversion}(b).
The existence of tunneling channels requires $e\mathcal{E}L\ge\Delta'(k_y)=\sqrt{\mu^2_{ex}(k_y)+\Delta^2}$, which gives an upper bound for $k_y^2$,
\begin{equation}
  k_y^2\le k_{y,max}^2= m[\mu^0_{ex}+\sqrt{(e\mathcal{E}L)^2-\Delta^2}].
\end{equation}

In this case, the classical turning points are $x'_{\pm}=(\pm\Delta'/2- E)/e\mathcal{E}$.
In addition to the region $x_{-}\le x\le x_{+}$, the complex wavevector $k(x)$ also has an imaginary part in the region $x'_{-}\le x\le x_{-}$ and $x_{+}\le x\le x'_{+}$ which is
\begin{equation}
  |\mathrm{Im}k(x)|=\sqrt{m}
    \sqrt{|\mu_{ex}|-\sqrt{4(E+e\mathcal{E}x)^2-\Delta^2}}.\nonumber
\end{equation}
The Zener parameter in this case is 
\begin{align}
  \zeta(k_y)
  =&\frac{\sqrt{m}\Delta^{3/2}}{\sqrt{2}e\mathcal{E}}\bigg[\int_0^{1}\rd\varepsilon\;\sqrt{\sqrt{\tilde{\mu}_{ex}^2+1-\varepsilon^2}+|\tilde{\mu}_{ex}|}\nonumber\\
  &+\sqrt{2}\int_{1}^{\sqrt{1+\tilde{\mu}^2_{ex}}}\rd\varepsilon\;\sqrt{|\tilde{\mu}_{ex}|-\sqrt{\varepsilon^2-1}}\bigg],\label{eq:zener_no_band_invert}
\end{align}
where $\tilde{\mu}_{ex}=\mu_{ex}(k_y)/\Delta=(\mu^0_{ex}-k_y^2/m)/\Delta<0$.
Then the WKB tunneling probability is $P(k_y)=\re^{-2\zeta(k_y)}$ and the current density is 
\begin{equation}
  j(k_y)=-\frac{eP(k_y)}{2\pi}\delta E(k_y),\label{eq:current_no_band_invert}
\end{equation}
where $\delta E(k_y)=e\mathcal{E}L-\Delta'$.

Combine Eq. \eqref{eq:current_band_invert}\eqref{eq:current_no_band_invert} and integration over $k_y$ gives the finally expression for the tunneling current density
\begin{align}
  j=-\frac{e}{(2\pi)^2}\int_{-k_{y,max}}^{k_{y,max}}\rd k_y\; \re^{-2\zeta(k_y)}\delta E(k_y),
  \label{eq:tunneling_current}
\end{align}
where $\delta E(k_y)=\min(2(e\mathcal{E}L-\Delta),\Delta'-\Delta)$ for $k_y^2\le m\mu_{ex}^0$ and is $e\mathcal{E}L-\Delta'$ for $k_y^2> m\mu_{ex}^0$
Besides, the Zener parameter $\zeta(k_y)$ is given by Eq. \eqref{eq:zener_band_invert} for $k_y^2 \le m\mu^0_{ex}$ and is Eq. \eqref{eq:zener_no_band_invert} for $k_y^2> m\mu^0_{ex}$.

The integration in Eq. \eqref{eq:tunneling_current} could not be solved analytically, but we can give an upper estimation for the tunneling current.
The Zener parameter $\zeta(k_y)$ is a monotonically increasing function of $k_y^2$, thus 
\begin{align}
  \zeta\ge\frac{\sqrt{m}\Delta^{3/2}}{\sqrt{2}e\mathcal{E}}\int_0^{1}\rd \varepsilon\;\sqrt{\sqrt{(\tilde{\mu}^0_{ex})^2+1-\varepsilon^2}-\tilde{\mu}^0_{ex}},
\end{align}
where $\tilde{\mu}^0_{ex}=\mu^0_{ex}/\Delta$.
It's convenient to define the correlation length of the gap (penetration depth of the band electron wavefunction into the classically forbidden region)
\begin{equation}
  \xi^{-1}=\sqrt{2m\Delta}\int_0^{1}\rd \varepsilon\;\sqrt{\sqrt{(\tilde{\mu}^0_{ex})^2+1-\varepsilon^2}-\tilde{\mu}^0_{ex}}
\end{equation}
and the tunneling length $\ell\equiv \Delta/e\mathcal{E}$.
Then the tunneling probability is approximated as $P=\re^{-2\zeta}\le \re^{-\frac{\ell}{\xi}}$.
Besides, one could verify that $\delta E(k_y)\le 2(e\mathcal{E}L-\Delta)$, so an upper bound for the current density is estimated as
\begin{align}
  |j|<&\frac{2e(e\mathcal{E}L-\Delta)\re^{-\frac{\ell}{\xi}}}{(2\pi)^2}\int_{-k_{y,max}}^{k_{y,max}}\rd k_y\nonumber\\
  =&\frac{e(e\mathcal{E}L-\Delta)\re^{-\frac{\ell}{\xi}}}{\pi^2}\sqrt{m[\mu^0_{ex}+\sqrt{(e\mathcal{E}L)^2-\Delta^2}]},\nonumber
\end{align}
which generates a Zener tunneling current in the form of
\begin{equation}
  I_{z}\sim (e\mathcal{E}L-\Delta)^{3/2}\re^{-\frac{\ell}{\xi}}
\end{equation}
in the thermodynamic limit $e\mathcal{E}L\gg \Delta$.

In excitonic insulators, $\mu^0_{ex}$ appeared in this section should be understood as the exciton chemical potential normalized by original band gap $E_g$ and the HF self-energy $\Sigma^{HF}$, i.e $\mu'_{ex}=\mu^0_{ex}-E_g-\mathrm{Tr}(\Sigma^{HF}\sigma_z)$. 
And the normalized exciton chemical potential $\mu'_{ex}$ is roughly related with exciton density as $\mu'_{ex}=k_F^2/m=4\pi n_{ex}/m$.
Then the correlation length as a function of exciton density is
\begin{align}
  \xi^{-1}
  =&\sqrt{2m\Delta}\int_0^1\rd\varepsilon\;\sqrt{\sqrt{(\mu'_{ex})^2/\Delta^2+1-\varepsilon^2}-{\mu'_{ex}}/{\Delta}}\nonumber\\
  =&\frac{m\Delta}{\sqrt{2\pi n_{ex}}}\int_0^{1}\rd \varepsilon\;\frac{\sqrt{1-\varepsilon^2}}{\sqrt{\sqrt{1+(1-\varepsilon^2)(m\Delta/4\pi n_{ex})^2}+1}}.\label{eq:correlation_length}
\end{align}
In the high exciton density limit
\begin{equation}
  \xi^{-1}\approx \frac{m\Delta}{\sqrt{2\pi n_{ex}}}\frac{\int_0^1\rd\varepsilon\;\sqrt{1-\varepsilon^2}}{\sqrt{2}}=\frac{\pi m\Delta}{8\sqrt{\pi n_{ex}}}=\frac{\pi\Delta}{4v_F}\label{eq:correlation_length_high_density}
\end{equation}
is just the coherence length of the excitonic insulator.

\bibliography{ref.bib}
\bibliographystyle{apsrev4-2}


\end{document}